\documentclass[pra,preprint,preprintnumbers,amsmath,amssymb,nofootinbib]{revtex4}
\usepackage{graphicx}
\usepackage{bm}
\begin{document}

\title{Entanglement balance of quantum $(e,2e)$ scattering processes}

\author{Konstantin A. Kouzakov}
\affiliation{Department of Nuclear Physics and Quantum Theory of Collisions, Faculty of Physics,
Lomonosov Moscow State University, Moscow 119991, Russia}
\author{Levan Chotorlishvili}
\affiliation{Institut f\"ur Physik, Martin-Luther Universit\"at Halle-Wittenberg, 06099 Halle/Saale, Germany}
\author{Jonas W\"atzel}
\affiliation{Institut f\"ur Physik, Martin-Luther Universit\"at Halle-Wittenberg, 06099 Halle/Saale, Germany}
\author{Jamal Berakdar}
\affiliation{Institut f\"ur Physik, Martin-Luther Universit\"at Halle-Wittenberg, 06099 Halle/Saale, Germany}
\author{Arthur Ernst}
\affiliation{Institute for Theoretical Physics, Johannes Kepler University, Altenberger Strasse 69, 4040 Linz, Austria\\
and Max-Planck-Institut f\"ur Mikrostrukturphysik, Weinberg 2, D-06120 Halle, Germany}
%

\begin{abstract}
The theory of quantum information constitutes the functional value of the quantum entanglement, i.e., quantum entanglement is essential for high fidelity of quantum protocols, while fundamental physical processes behind the formation of quantum entanglement are less relevant for practical purposes. In the present work, we explore physical mechanisms leading to the emergence of quantum entanglement in the initially disentangled system. In particular, we analyze spin entanglement of outgoing electrons in a nonrelativistic quantum $(e,2e)$ collision on a target with one active electron. Our description exploits the time-dependent scattering formalism for typical conditions of scattering experiments, and contrary to the customary stationary formalism operates with realistic scattering states. We quantify the spin entanglement in the final scattering channel through the pair concurrence and express it in terms of the experimentally measurable spin-resolved $(e,2e)$ triple differential cross sections. Besides, we consider Bell's inequality and inspect the regimes of its violation in the final channel. We address both the pure and the mixed initial spin state cases and uncover kinematical conditions of the maximal entanglement of the outgoing electron pair. The numerical results for the pair concurrence, entanglement of formation, and violation of Bell's inequality obtained for the $(e,2e)$ ionization process of atomic hydrogen show that the entangled electron pairs indeed can be formed in the $(e,2e)$ collisions even with spin-unpolarized projectile and target electrons in the initial channel. The positive entanglement balance---the difference between entanglements of the initial and final electron pairs---can be measured in the experiment.
\end{abstract}

\maketitle

%
\section{Introduction}
\label{intro}
The quantum entanglement of particles in a particle pair represents one of the marked phenomena that falls beyond the scope of the conventional paradigm of classical physics~\cite{EPR}. It stems from the impossibility of assigning specific properties to one constituent of the pair being in an entangled quantum state. The famous example is the singlet state of two spin-1/2 particles~\cite{Bohm1957} termed as a Bell state $\Psi^-_{\rm Bell}$. Whatever the distance between particles is, their spin states remain inherently correlated in such a way that the measurement of the spin state of the first particle determines the spin state of the second particle.
This feature owes to the nonseparability of the pair wave function $\Psi^-_{\rm Bell}$ and reflects the intrinsic nonlocality of quantum mechanics as opposed to its classical counterpart relying on local realism.
Beyond its fundamental significance, the concept of quantum entanglement lies in the basement of quantum information theory and quantum computing~\cite{suter, Braunstein, HorodeckiRMP, Werner, Zurek1, Osterloh, Cirac, Plenio1, Nori1, Nori2, Plenio2, Das Sarma, Sanders, Adesso, A. del Campo, Wilde, Kohler}. Quantum entanglement is a natural ingredient for any communication channel connecting futuristic quantum engines and devises. Harvesting, storage or manipulation of quantum information requires a substantive degree of quantum entanglement. The incisive issue is either quantification of quantum entanglement or minimization of adverse environmental effects. In the literature, quantum entanglement is mainly considered as a tool \cite{Zueco1, W. H. Zurek, D. Loss, Lutz, Levan1, Levan2, Levan3, Bose, Nori3, Azimi, Zueco2, Lukin}. 
Nevertheless, the problem of vital interest concerns quantum mechanical processes leading to the emergence of quantum entanglement in an initially disentangled system. Therefore, in the present work, we aim at revealing theoretically the mechanisms responsible for the formation of spin entanglement of two identical fermions, specifically of an outgoing electron pair emerged in an ionizing electron-target collision. We also propose an experimentally feasible protocol for measuring the spin entanglement in $(e,2e)$ scattering processes~\cite{Whelan1994,Coplan_RMP1994}. 

A simple picture of the quantum $(e,2e)$ scattering process involves the initial and final asymptotic states. The projectile and target electrons in the initial asymptotic state before the collision are far apart. The projectile impinges on the target so that the particles interact and diverge from each other, approaching the final asymptotic state when particles again are far apart.  Suppose that the initial electron-pair state is not entangled, and the final state is. Then one could characterize the quantum $(e,2e)$ scattering process by a positive entanglement balance. Usually, the details of the interparticle interactions during the collision are inaccessible, and one is only able to measure the probability of the transition between the initial and final asymptotic states. The basic question that we address in this study is whether or not such information suffices for signaling the formation of quantum entanglement as a result of the collision.

It is well known (see, for instance, the textbooks~\cite{Taylor_book, Goldberger_book}) that in the free electron-electron scattering the differential cross sections for the singlet and triplet electron pairs are not the same. The reason is the difference between the symmetric and asymmetric electron-pair spatial wave functions. In particular, the triplet cross section vanishes in the symmetric kinematics, i.e., at equal energy sharing, while the singlet one remains finite. This feature can be used~\cite{Lamata2006} for selecting singlet electron pairs in free electron-electron collisions, thereby creating a maximally entangled Bell state $\Psi^-_{\rm Bell}$ of two electrons in the final scattering channel. One might expect that the singlet electron pairs can be also selected in electron-electron collisions where one of the electrons is initially bound (for example, bound to the solid surface~\cite{Feder2015,Vasilyev2017}). It should be emphasized that the spin entanglement is formed as a result of the scattering process in the system of two electrons which are initially disentangled. This fact cannot be properly accounted for within the time-independent scattering formalism, in which one describes the electrons in the initial and final asymptotic states with non-square-integrable, spatially delocalized wave functions (plane waves) that are not spatially separated.     
The spatial indistinguishability of particles leads to the spurious entanglement \cite{Schliemann2001, Ghirardi2004}. This forces one~\cite{Lamata2006} to modify the standard criteria of entanglement, for example, such as the von Neumann entropy of the reduced density matrix of the pair state. Another important issue that one should address in the theoretical treatment of spin entanglement formed in the scattering process is the case of a mixed initial pair spin state, i.e., when prior the collision the projectile and target electrons are only partially polarized or even unpolarized. Indeed, such a case is most typical for scattering experiments, where the preparation of a pure spin-polarized initial electron-pair state constitutes practically an intractable task (see, for instance, a very useful monograph of Kessler~\cite{Kessler_monograph}). This means, in particular, that the modified von Neumann entropy usually employed as the entanglement measure is not valid in the interpretation and analysis of the data of the scattering experiments since this measure is generally valid only for pure pair states.

In the present work, we develop the comprehensible theoretical formulation avoiding the drawbacks indicated above. First, we employ the time-dependent scattering theory~\cite{Taylor_book}, which operates with square-integrable, spatially localized electron wave functions (wave packets) in the initial and final scattering channels. In this way we have spatially separated projectile and target electrons before and after the collision and, accordingly, no spurious entanglement can arise provided one performs local measurements of the electrons. Second, to quantify the entanglement balance of the $(e,2e)$ scattering process, we utilize such criteria as the pair concurrence and entanglement of formation~\cite{Wootters}. In contrast to the von Neumann entropy of the reduced density matrix of the pair state, the pair concurrence and entanglement of formation are applicable for both the pure and mixed pair states. Note that due to the spatial separation of the electrons in our approach they can be treated as distinguishable particles~\cite{Tichy2011, Tichy2013, Franco2016} and therefore no modification of the pair concurrence and entanglement of formation is needed.

The paper is structured as follows. In Sec.~\ref{gen-form} we deliver the general theory for an elementary $(e,2e)$ collision and derive the spin state of the final electron pair. The entanglement measures and Bell's inequality for quantifying the entanglement balance of the studied quantum scattering process are formulated in Sec.~\ref{entangle}. Then, in Sec.~\ref{Spin_avg}, we take into account the ensemble average over impact parameters and spin states of the colliding electron pairs in a scattering experiment. Sec.~\ref{scat_amplitude} is devoted to the general symmetry properties of the $(e,2e)$ scattering amplitudes. In Sec.~\ref{res} we present and discuss numerical results for the entanglement measures and Bell's inequality violation in the case of $(e,2e)$ ionization of atomic hydrogen. Sec.~\ref{concl} summarizes this work. Atomic units, $e=\hbar=m_e=1$, are used throughout unless otherwise specified.

\section{General formulation}
\label{gen-form}
We consider the process where an electron with momentum ${\bf k}_0$ impinges on a target ${\mathcal T}$ with one active electron (e.g., atomic hydrogen) and induces the $(e,2e)$ collision
\begin{equation}
\label{(e,2e)_collision}
e^- +{\mathcal T}\to\mathcal{T}^+
+2e^-.
\end{equation}
As a result, two outgoing electrons (scattered and ejected) are emerged having asymptotic momenta ${\bf k}_A$ and ${\bf k}_B$. In what follows, we assume the target to be infinitely heavy and at rest, so that the center-of-mass and laboratory frames of reference coincide.

We focus on the analysis of spin entanglement in the outgoing electron pair. Therefore, we should examine how the initial spin state of the projectile-target system changes due to the $(e,2e)$ collision~(\ref{(e,2e)_collision}). First, we construct the so-called in asymptote $|\Psi_{\rm in}\rangle$~\cite{Taylor_book} of the projectile-target system accounting for electron spins and, for the moment, treating electrons as distinguishable particles. Let the spin states of the ingoing and target electrons be given by, respectively,
\begin{equation}
\label{one-electron_spin_states}
|\chi(1)\rangle=\alpha|1\uparrow\rangle+\beta|1\downarrow\rangle, \qquad
|\eta(2)\rangle=\gamma|2\uparrow\rangle+\delta|2\downarrow\rangle,
\end{equation}
where $|\uparrow\rangle$ ($|\downarrow\rangle$) is a spin-up (spin-down) spinor, and $|\alpha|^2+|\beta|^2=|\gamma|^2+|\delta|^2=1$. The two-electron in asymptote can be presented as~\cite{Taylor_book,Goldberger_book}
\begin{equation}
\label{in-asymptote}
|\Psi_{\rm in}(1,2)\rangle=
|\psi_{{\bf k}_0}^{({\bf b})}(1)\chi(1)\rangle\otimes|\psi_{\mathcal T}(2)\eta(2)\rangle,
\end{equation}
where $\psi_{{\bf k}_0}^{({\bf b})}$ is the projectile wave packet displaced by the impact parameter ${\bf b}$ from the axis directed along ${\bf k}_0$ and intersecting the center of the target, according to $\langle{\bf p}|\psi_{{\bf k}_0}^{({\bf b})}\rangle=\phi_{{\bf k}_0}^{({\bf b})}({\bf p})=e^{-i{\bf pb}}\phi_{{\bf k}_0}({\bf p})$, with the momentum-space wave function $\phi_{{\bf k}_0}({\bf p})$ peaked about
${\bf k}_0$, and $\psi_{\mathcal T}$ is a bound state of the target. The state~(\ref{in-asymptote}) is the asymptotic state of the projectile-target system before the collision, namely as $t\to-\infty$, in the interaction representation. In the Schr\"odinger representation it is given by
\begin{equation}
\label{in-asymptote_t=-oo}
|\Psi(1,2)\rangle_{t=-\infty}=\lim_{t\to-\infty}\hat{U}_0^{(1)}(t,0)|\Psi_{\rm in}(1,2)\rangle
=|\psi_{{\bf k}_0}^{({\bf b})}(1,t=-\infty)\chi(1)\rangle\otimes|\psi_{\mathcal T}(2,t=-\infty)\eta(2)\rangle,
\end{equation}
with the evolution operator
$$
\hat{U}^{(1)}_0(t,t')=e^{-i\hat{H}^{(1)}_0(t-t')}, \qquad \hat{H}^{(1)}_0=\hat{H}-\hat{V}_{1{\mathcal T}},
$$
where $\hat{H}$ is the full Hamiltonian of the projectile-target system and $\hat{V}_{1{\mathcal T}}$ is the projectile-target interaction. Now let us take into account the indistinguishability of electrons by antisymmetrizing the wave function of the system.
Acting with the antisymmetrization operator $\hat{\Lambda}_a$
on Eq.~(\ref{in-asymptote_t=-oo}) yields
\begin{eqnarray}
\label{in-asymptote_t=-oo_a}
\hat{\Lambda}_a|\Psi(1,2)\rangle_{t=-\infty}&=&|\Psi^{(a)}(1,2)\rangle_{t=-\infty}\nonumber\\
&=&\frac{1}{\sqrt{2}}\left[|\psi_{{\bf k}_0}^{({\bf b})}(1,t=-\infty)\chi(1)\rangle\otimes|\psi_{\mathcal T}(2,t=-\infty)\eta(2)\rangle \right. \nonumber\\
&{}&\left.-|\psi_{\mathcal T}(1,t=-\infty)\eta(1)\rangle\otimes|\psi_{{\bf k}_0}^{({\bf b})}(2,t=-\infty)\chi(2)\rangle\right].
\end{eqnarray}
This is a Slater determinant of the two one-electron states which are orthogonal due to $\langle\psi_{{\bf k}_0}^{({\bf b})}(t=-\infty)|\psi_{\mathcal T}(t=-\infty)\rangle=0$ that stems from the spatial non-overlap of the projectile wave packet with the target wave function in the asymptotic limit $t\to-\infty$. Therefore, though it cannot be expressed as a product of one-particle states, the state~(\ref{in-asymptote_t=-oo_a}) describes two independent spatially separated electrons and, hence, is not genuinely entangled (see, for instance, Ref.~\cite{Franco2016} and references therein). This means that a local measurement of a property of one electron, for example, such as spin, will always yield the state $|\chi\rangle$ ($|\eta\rangle$) if the electron is located at $|\psi_{{\bf k}_0}^{({\bf b})}\rangle$ ($|\psi_{\mathcal T}\rangle$)~\cite{Tichy2011}. 
The state of the system  at any given moment $t_0$ is
\begin{eqnarray}
\label{phys_state}
|\Psi^{(a)}(1,2)\rangle_{t_0}=\hat{U}(t_0,-\infty)|\Psi^{(a)}(1,2)\rangle_{t=-\infty}, \qquad \hat{U}(t,t')=e^{-i\hat{H}(t-t')}.
\end{eqnarray}
At $t_0=\infty$ it contains asymptotic components corresponding to the final channels of the possible scattering processes in the system, including the $(e,2e)$ collision~(\ref{(e,2e)_collision}) and the elastic and inelastic scattering processes
$$
e^-+{\mathcal T}\to{\mathcal T}+e^-, \qquad e^-+{\mathcal T}\to{\mathcal T}^*+e^-.
$$
The $(e,2e)$ asymptotic component is determined by
\begin{equation}
\label{(e,2e)_asymptote_final_ch}
|\Psi_{(e,2e)}^{(a)}(1,2)\rangle_{t=\infty}=\lim_{t\to\infty}\hat{U}_0(t,0)|\Psi_{\rm out}^{(e,2e)}(1,2)\rangle, \qquad
\hat{U}_0(t,t')=e^{-i\hat{H}_0(t-t')},
\end{equation}
where $\hat{H}_0=\hat{H}-\hat{V}_{{1\mathcal T}^+}-\hat{V}_{{2\mathcal T}^+}-\hat{V}_{12}$ is the asymptotic free Hamiltonian\footnote{In general, the potentials $\hat{V}_{{1\mathcal T}^+}$, $\hat{V}_{{2\mathcal T}^+}$, and $\hat{V}_{12}$ are long range, so that the Hamiltonian $\hat{H}$ does not become free, i.e., $\hat{H}_0$, in the discussed asymptotic limit. This implies that the usual formalism of the multichannel scattering theory developed for short-range potentials is not directly applicable. The difficulty is circumvented by properly modifying the $S$ and $T$ matrices~\cite{Shablov2010}.} in the final channel of the process~(\ref{(e,2e)_collision}). The $(e,2e)$ out asymptote is given by
\begin{equation}
\label{e2e_out_asymptote}
|\Psi_{\rm out}^{(e,2e)}(1,2)\rangle=\frac{1}{\sqrt{2}}\left[\hat{S}_{(e,2e)}{(1,2)}|\Psi_{\rm in}(1,2)\rangle-
\hat{S}_{(e,2e)}{(2,1)}|\Psi_{\rm in}(2,1)\rangle\right],
\end{equation}
where $\hat{S}_{(e,2e)}{(1,2)}$ and $\hat{S}_{(e,2e)}{(2,1)}$  are the scattering operators for the $(e,2e)$ transition in the case of distinguishable electrons~\cite{Goldberger_book}:
\begin{subequations}
\label{S-operators}
\begin{align}
 \hat{S}_{(e,2e)}{(1,2)}&=\hat{U}_0^\dag(\infty,0)\hat{U}(\infty,-\infty)\hat{U}_0^{(1)}(-\infty,0), \\
\hat{S}_{(e,2e)}{(2,1)}&=\hat{U}_0^\dag(\infty,0)\hat{U}(\infty,-\infty)\hat{U}_0^{(2)}(-\infty,0).
\end{align}
\end{subequations}
The state~(\ref{(e,2e)_asymptote_final_ch}) describes two outgoing electrons that propagate freely at asymptotically large distances from the collision region with their spatial wave functions being well separated.

For obtaining the spin state $|X_f\rangle$ of the outgoing electron pair in the final channel of the process~(\ref{(e,2e)_collision})  we must project the $(e,2e)$ asymptotic component~(\ref{(e,2e)_asymptote_final_ch}) onto the two-electron plane-wave state
$|{\bf k}_A,{\bf k}_B\rangle=|{\bf k}_A\rangle\otimes|{\bf k}_B\rangle$. The indicated projection mimics the typical coincident measurement of the electron energies and angles in two spatially separated, distinct detectors $A$ (Alice) and $B$ (Bob).
Using Eqs.~(\ref{(e,2e)_asymptote_final_ch}) and~(\ref{e2e_out_asymptote}), we get for the (unnormalized) final spinor
\begin{eqnarray}
\label{final_spin_state}
|X_f\rangle&=&\langle{\bf k}_A,{\bf k}_B|\Psi_{(e,2e)}^{(a)}(1,2)\rangle_{t=\infty}\nonumber\\
&=&\frac{1}{\sqrt{2}}\left\{\lim_{t\to\infty}\exp\left[-i\left(\frac{k_A^2}{2}+\frac{k_B^2}{2}\right)t\right]\right\}
\left[\left(\langle{\bf k}_A,{\bf k}_B|\hat{S}_{(e,2e)}(1,2)|\psi_{{\bf k}_0}^{({\bf b})},\psi_{\mathcal T}\rangle\right)|\chi(1)\rangle\otimes|\eta(2)\rangle\right. \nonumber\\
&{}&\left.-\left(\langle{\bf k}_A,{\bf k}_B|\hat{S}_{(e,2e)}(2,1)|\psi_{\mathcal T},\psi_{{\bf k}_0}^{({\bf b})}\rangle\right)|\eta(1)\rangle\otimes|\chi(2)\rangle\right].
\end{eqnarray}
Here it is taken into account that in the discussed nonrelativistic case the interactions in the colliding system are spin independent and, hence, the scattering operator does not act on the spin states. The $S$-matrix elements can be expressed in terms of the $T$ matrix~\cite{Taylor_book} as
\begin{eqnarray}
\langle{\bf k}_A,{\bf k}_B|\hat{S}_{(e,2e)}(1,2)|\psi_{{\bf k}_0}^{({\bf b})},\psi_{\mathcal T}\rangle&=&-i\int\frac{d^3 p}{(2\pi)^2}\,t({\bf p},\psi_{\mathcal T}\to{\bf k}_A,{\bf k}_B)e^{-i{\bf pb}}\phi_{{\bf k}_0}({\bf p})\nonumber\\
&{}&\times \delta\left(\frac{p^2}{2}+E_{\mathcal T}-E_A-E_B\right),
\label{non-spin_part_a_projection}
\\
\label{non-spin_part_s_projection}
\langle{\bf k}_A,{\bf k}_B|\hat{S}_{(e,2e)}(2,1)|\psi_{\mathcal T},\psi_{{\bf k}_0}^{({\bf b})}\rangle&=&-i\int\frac{d^3 p}{(2\pi)^2}\,t({\bf p},\psi_{\mathcal T}\to{\bf k}_B,{\bf k}_A)e^{-i{\bf pb}}\phi_{{\bf k}_0}({\bf p})\nonumber\\
&{}&\times \delta\left(\frac{p^2}{2}+E_{\mathcal T}-E_A-E_B\right),
\end{eqnarray}
where $E_{A,B}=k_{A,B}^2/2$ are the energies of the outgoing electrons and $E_{\mathcal T}$ is the energy of the target electron state. If the region in the vicinity of  ${\bf k}_0$ where $\phi_{{\bf k}_0}({\bf p})$ is appreciably different from zero is so small that the variation of the $T$ matrices in this region is insignificant, one can replace~\cite{Taylor_book} their values in the integrands of Eqs.~(\ref{non-spin_part_a_projection}) and~(\ref{non-spin_part_s_projection}) by those at ${\bf p}={\bf k}_0$. One thus obtains
\begin{eqnarray}
\langle{\bf k}_A,{\bf k}_B|\hat{S}_{(e,2e)}(1,2)|\psi_{{\bf k}_0}^{({\bf b})},\psi_{\mathcal T}\rangle&=&t_d\mathcal{F}({\bf k}_0,{\bf b};E_A,E_B,E_{\mathcal T}),
\label{non-spin_part_a_projection_2}
\\
\label{non-spin_part_s_projection_2}
\langle{\bf k}_A,{\bf k}_B|\hat{S}_{(e,2e)}(2,1)|\psi_{\mathcal T},\psi_{{\bf k}_0}^{({\bf b})}\rangle&=&t_e\mathcal{F}({\bf k}_0,{\bf b};E_A,E_B,E_{\mathcal T}),
\end{eqnarray}
where
\begin{equation}
\label{direct&exchange}
t_d=t({\bf k}_0,\psi_{\mathcal T}\to{\bf k}_A,{\bf k}_B) \qquad \text{and} \qquad t_e=t({\bf k}_0,\psi_{\mathcal T}\to{\bf k}_B,{\bf k}_A)
\end{equation}
are the on-shell $T$ matrices ($E_0+E_{\mathcal T}=E_A+E_B$, with $E_0=k_0^2/2$) called the direct and exchange $(e,2e)$ scattering amplitudes, and
\begin{eqnarray}
\label{F}
\mathcal{F}({\bf k}_0,{\bf b};E_A,E_B,E_{\mathcal T})&=&-i\int\frac{d^3 p}{(2\pi)^2}\,e^{-i{\bf pb}}\phi_{{\bf k}_0}({\bf p})\delta\left(\frac{p^2}{2}+E_{\mathcal T}-E_A-E_B\right).
\end{eqnarray}
Taking into account Eqs.~(\ref{non-spin_part_a_projection_2}) and~(\ref{non-spin_part_s_projection_2}), from Eq.~(\ref{final_spin_state}) we deduce that 
\begin{eqnarray}
\label{final_spin_state_1}
|X_f\rangle&=&\frac{1}{\sqrt{2}}\left\{\lim_{t\to\infty}\exp\left[-i\left(\frac{k_A^2}{2}+\frac{k_B^2}{2}\right)t\right]\right\}\mathcal{F}({\bf k}_0,{\bf b};E_A,E_B,E_{\mathcal T})\nonumber\\
&{}&\times\left[t_d|\chi(1)\rangle\otimes|\eta(2)\rangle-t_e|\eta(1)\rangle\otimes|\chi(2)\rangle\right].
\end{eqnarray}
Using the Bell states
\begin{eqnarray}
\label{Bell_states}
|\Phi^{\pm}_{\rm Bell}\rangle=\frac{1}{\sqrt{2}}(|1\uparrow\rangle\otimes|2\uparrow\rangle\pm|1\downarrow\rangle\otimes|2\downarrow\rangle), \qquad
|\Psi^{\pm}_{\rm Bell}\rangle=\frac{1}{\sqrt{2}}(|1\uparrow\rangle\otimes|2\downarrow\rangle\pm|1\downarrow\rangle\otimes|2\uparrow\rangle),\nonumber\\
\end{eqnarray}
we can recast Eq.~(\ref{final_spin_state_1}) into the form
\begin{eqnarray}
\label{out-asymptote_2}
|X_{f}\rangle
&\propto&
(t_d-t_e)\left[(\alpha\gamma+\beta\delta)|\Phi^{+}_{\rm Bell}\rangle+(\alpha\gamma-\beta\delta)|\Phi^{-}_{\rm Bell}\rangle+(\alpha\delta+\beta\gamma)|\Psi^{+}_{\rm Bell}\rangle\right]
\nonumber\\
&{}&
+(t_d+t_e)(\alpha\delta-\beta\gamma)|\Psi^{-}_{\rm Bell}\rangle.
\end{eqnarray}
From Eq.~(\ref{out-asymptote_2}) it follows that the properties of the final spin state of the electron pair depend on the values of the direct $t_d$ and exchange $t_e$ scattering amplitudes. In particular, when $t_d=t_e$ this state becomes a completely entangled Bell's state $\Psi^-_{\rm Bell}$, which is known to maximally violate Bell's inequality. For addressing the issue of quantification of entanglement of the pair spinor $|X_f\rangle$ in the next section, we shall need the normalized density matrix of the state~(\ref{final_spin_state_1}):
\begin{eqnarray}
\label{rho_f}
\hat{\rho}_f&=&\frac{|X_f\rangle\langle X_f|}{||X_f||^2} \nonumber\\
&=&\frac{1}{u}\left[|t_d|^2(|\chi(1)\rangle\langle\chi(1)|)\otimes(|\eta(2)\rangle\langle\eta(2)|)
+|t_e|^2(|\eta(1)\rangle\langle\eta(1)|)\otimes(|\chi(2)\rangle\langle\chi(2)|)\right.\nonumber\\
&{}&\left.-t_dt_e^*(|\chi(1)\rangle\langle\eta(1)|)\otimes(|\eta(2)\rangle\langle\chi(2)|)
-t_d^*t_e(|\eta(1)\rangle\langle\chi(1)|)\otimes(|\chi(2)\rangle\langle\eta(2)|)\right],
\end{eqnarray}
where
\begin{equation}
\label{normalization_spin_wf}
{u}=|t_d|^2+|t_e|^2-2{\rm Re}(t_dt_e^*)|\alpha\gamma^*+\beta\delta^*|^2=|t_d-t_e|^2+2{\rm Re}(t_dt_e^*)
|\alpha\delta-\beta\gamma|^2.
\end{equation}
We can interpret Eq.~(\ref{rho_f}) as a state of two effectively distinguishable electrons, since it describes the spin state of electrons 1 and 2 which are measured by two distinct detectors $A$ and $B$~\cite{Tichy2013}, respectively. Let us also introduce the following unnormalized density matrix:
\begin{equation}
\label{rho_f_unnorm}
\hat{\tilde\rho}_f=\int\limits_0^\infty dE_{A}\int d^2b\,\frac{2k_Ak_B}{(2\pi)^6}\,|X_f\rangle\langle X_f|=
\frac{k_B}{(2\pi)^6}
\int\limits_0^\infty dE_{A}\,k_Au\hat{\rho}_f\int d^2b\,|\mathcal{F}({\bf k}_0,{\bf b};E_A,E_B,E_{\mathcal T})|^2.
\end{equation}
The integration in the case of the ingoing wave packet $\phi_{{\bf k}_0}({\bf p})$ sharply peaked about ${\bf k}_0$ is straightforward (see Appendix~\ref{integral_wave_packet}), yielding
\begin{equation}
\label{rho_f_unnorm_1}
\hat{\tilde\rho}_f=\frac{k_Ak_Bu}{(2\pi)^5k_0}\,\hat{\rho}_f,
\end{equation}
where $E_A=E_0+E_{\mathcal T}-E_B$. Using the density matrix~(\ref{rho_f_unnorm_1}) one can derive~\cite{Goldberger_book} the triple differential cross section (TDCS) of the $(e,2e)$ scattering process. For the TDCS in the case of the $(e,2e)$ transition to the spin state $|\chi_A(1)\rangle\otimes|\eta_B(2)\rangle$ one has
\begin{equation}
\label{TDCS_spin-resolved}
\frac{d\sigma_{\chi\eta\to\chi_A\eta_B}}{dE_Bd\Omega_Ad\Omega_B}=
{\rm Tr}\left(\hat{\tilde\rho}_f(|\chi_A(1)\rangle\langle\chi_A(1)|)\otimes(|\eta_B(2)\rangle\langle\eta_B(2)|)\right),
\end{equation}
where $\Omega_A$ and $\Omega_B$ specify solid angles of the outgoing electrons.
In particular, the spin-unresolved TDCS is given by
\begin{equation}
\label{TDCS_spin-unresolved}
\frac{d\sigma_{\chi\eta}}{dE_Bd\Omega_Ad\Omega_B}=
{\rm Tr}\hat{\tilde\rho}_f=\frac{k_Ak_Bu}{(2\pi)^5k_0}.
\end{equation}
From Eq.~(\ref{TDCS_spin-resolved}) the
following basic results~\cite{Joachain_book} can be derived:
\begin{subequations}
\label{TDCS_spin-resolved_basic}
\begin{align}
I_{\uparrow\uparrow}&=\frac{d\sigma_{\uparrow\uparrow\to\uparrow\uparrow}}{dE_Bd\Omega_Ad\Omega_B}=\frac{d\sigma_{\downarrow\downarrow\to\downarrow\downarrow}}{dE_Bd\Omega_Ad\Omega_B}=\frac{k_Ak_B}{(2\pi)^5k_0}\,|t_d-t_e|^2, \\
I_{\uparrow\downarrow}^{(d)}&=\frac{d\sigma_{\uparrow\downarrow\to\uparrow\downarrow}}{dE_Bd\Omega_Ad\Omega_B}=\frac{d\sigma_{\downarrow\uparrow\to\downarrow\uparrow}}{dE_Bd\Omega_Ad\Omega_B}=
\frac{k_Ak_B}{(2\pi)^5k_0}\,|t_d|^2, \\
 I_{\uparrow\downarrow}^{(e)}&=\frac{d\sigma_{\uparrow\downarrow\to\downarrow\uparrow}}{dE_Bd\Omega_Ad\Omega_B}=\frac{d\sigma_{\downarrow\uparrow\to\uparrow\downarrow}}{dE_Bd\Omega_Ad\Omega_B}=\frac{k_Ak_B}{(2\pi)^5k_0}\,|t_e|^2.
\end{align}
\end{subequations}
Since for all other spin transitions the TDCS is zero, the quantities $I_{\uparrow\uparrow}$ and $I_{\uparrow\downarrow}=I^{(d)}_{\uparrow\downarrow}+I^{(e)}_{\uparrow\downarrow}$ amount to the spin-unresolved TDCSs for $(e,2e)$ scattering with parallel and antiparallel electron spins, respectively.

\section{Entanglement criteria}
\label{entangle}
There are various entanglement witnesses either for pure or mixed bipartite quantum states $\hat{\rho}_{12}$~\cite{Mintert2005}. For a bipartite quantum system in the disentangled pure state, tracing out one of the parts, $\hat{\rho}_{1}={\rm Tr}_{2}\hat{\rho}_{12}$, leaves the system still in the pure state, that is, ${\rm Tr}\,\hat{\rho}_{1}^{2}=1$. In contrast, in the case of the entangled pure state, the reduced density matrix $\hat{\rho}_{1}$ appears to be always mixed, i.e., ${\rm Tr}\,\hat{\rho}_{1}^{2}<1$, and hence has a nonzero linear entropy $S_L=1-{\rm Tr}\,\hat{\rho}_{1}^{2}>0$, which is a lower approximation to the customary von Neumann entropy $S=-{\rm Tr}(\hat{\rho}_{1}\log_2\hat{\rho}_{1})\geq S_L$. Thus, if the bipartite quantum state is pure, mixedness of the reduced density matrix expressed in terms of the entropy measures is a valid entanglement witness. However, for mixed bipartite quantum states the nonzero entropy of the reduced density matrix is not a reliable criterion of entanglement anymore: one can get $S_L>0$ for a statistical mixture $\hat{\rho}_{12}=\sum_ip_i\hat{\rho}_{12}^{(i)}$, where $p_i>0$ and $\sum_ip_i=1$, of disentangled pure bipartite states $\hat{\rho}_{12}^{(i)}$, but it is not genuinely entangled.\footnote{A marked example is the disentangled mixed state $\hat{\rho}_{12}=\hat{\rho}_1\otimes\hat{\rho}_2$, with $\hat{\rho}_{1,2}=\frac{1}{2}\,\hat{I}$, for which one has $S_L=1/2$ and $S=1$.} A proper generalization of the entropy-based measure of entanglement that includes the case of mixed bipartite states is the entanglement of formation. The latter is typically calculated on the basis of the pair concurrence, which was originally introduced in Ref.~\cite{Wootters} as an auxiliary quantity but can be considered as an independent entanglement witness. Therefore, in order to explore the spin entanglement of the outgoing electron pair which can be both in a pure and in a mixed state, below we adopt the pair concurrence. In addition, we consider a violation of Bell's inequality.

\subsection{Entanglement measures}
As mentioned above, a frequently used measure of entanglement of a pure pair state is the von Neumann entropy of the reduced density matrix of this state. For the case of Eq.~(\ref{rho_f}) it is given by
\begin{equation}
\label{von_Neumann_entropy}
S_{f}=-{\rm Tr}(\hat{\rho}_{1,f}\log_2\hat{\rho}_{1,f}),
\end{equation}
where
\begin{eqnarray}
\label{post-scattering reduced1}
\hat{\rho}_{1,f}&=&{\rm Tr}_{2}\hat{\rho}_{f} \nonumber\\
&=&\frac{1}{u}\left[|t_d|^2|\chi(1)\rangle\langle\chi(1)|-t_dt_e^*(\alpha^*\gamma+\beta^*\delta)|\chi(1)\rangle\langle\eta(1)|-
t_d^*t_e(\alpha\gamma^*+\beta\delta^*)|\eta(1)\rangle\langle\chi(1)|\right.\nonumber\\
&{}&\left.+|t_e|^2|\eta(1)\rangle\langle\eta(1)|\right].
\end{eqnarray}
The entropy measure~(\ref{von_Neumann_entropy}) becomes in general inapplicable as a criterion of entanglement if we deal with a mixed pair state $\hat{\rho}=\sum_ip_i|\Psi_i\rangle\langle\Psi_i|$, where $p_i>0$. In the latter case one should rather use the entanglement of formation~\cite{Wootters}
\begin{equation}
\label{Entanglement_formation}
E_F(C_{\hat{\rho}})=h\left(\frac{1+\sqrt{1-C^2_{\hat{\rho}}}}{2}\right), \qquad h(x)=-x\log_2x-(1-x)\log_2(1-x),
\end{equation}
with the pair concurrence $C_{\hat{\rho}}$ defined as follows: $C_{\hat{\rho}}={\rm max}(0,\sqrt{\lambda_{1}}-\sqrt{\lambda_{2}}-\sqrt{\lambda_{3}}-\sqrt{\lambda_{4}}\big)$, with the eigenvalues $\lambda_{n=1,2,3,4}$, in decreasing order, of the matrix
$R=\hat{\rho}(\hat{\sigma}^{(1)}_{y}\otimes\hat{\sigma}^{(2)}_{y})\hat{\rho}^{*}(\hat{\sigma}^{(1)}_{y}\otimes \hat{\sigma}^{(2)}_{y})$. This definition of the concurrence is equivalent to
\begin{equation}
\label{conurrence_def}
C_{\hat{\rho}}=\underset{\{p_i,\Psi_i\}}{\rm inf}\sum_{i}p_iC_{\Psi_i},
\end{equation}
where $C_{\Psi_i}$ is the concurrence of the pure pair state $\Psi_i$ and the infinum is taken over all possible decompositions of $\hat{\rho}$ into pure states.

For a pure pair state such as~(\ref{rho_f}) the concurrence is given by~\cite{Rungta2001}
\begin{eqnarray}
\label{concurrence}
C_{f}=\sqrt{2(1-{\rm Tr}\,\hat{\rho}^{2}_{1,f})}
\end{eqnarray}
and the entanglement of formation~(\ref{Entanglement_formation}) amounts to the von Neumann entropy~(\ref{von_Neumann_entropy}). Inserting the reduced density matrix~(\ref{post-scattering reduced1}) into Eq.~(\ref{concurrence}), for the pair concurrence we obtain
\begin{eqnarray}
\label{concurrenceresult}
C_{f}=\frac{2}{u}|t_{d}||t_{e}||\alpha\delta-\beta\gamma|^{2}.
\end{eqnarray}
Both the concurrence~(\ref{concurrenceresult}) and the von Neumann entropy,
\begin{equation}
\label{von_Neumann_entropy1}
S_{f}=-\frac{1+\sqrt{1-C^2_{f}}}{2}\log_2\left(\frac{1+\sqrt{1-C^2_{f}}}{2}\right)-\frac{1-\sqrt{1-C_{f}^2}}{2}\log_2\left(\frac{1-\sqrt{1-C_{f}^2}}{2}\right),
\end{equation}
range from 0 to 1, with $C_{f}^{\rm min}=S_{f}^{\rm min}=0$ and $C_{f}^{\rm max}=S_{f}^{\rm max}=1$ corresponding to the disentangled and completely entangled pair states, respectively.

As was pointed out in the previous section, the pair spin state in the initial, pre-scattering channel is disentangled ($C_{i}=S_{i}=0$). From Eqs.~(\ref{concurrenceresult}) and~(\ref{von_Neumann_entropy1}) it follows that the pair spin state remains disentangled in the final, post-scattering channel if either of $t_d$, $t_{e}$, and $\alpha\delta-\beta\gamma$ equals zero. In contrast, when $t_d=t_e$ we have $C_{f}=1$ and $S_{f}=1$, so that the state is a maximally entangled Bell state $\Psi^-_{\rm Bell}$.

Let us adopt the Bloch-sphere representation of the spin states~(\ref{one-electron_spin_states}):
\begin{equation}
\label{Bloch_sphere}
\alpha=\cos\frac{\vartheta_{1}}{2}, \qquad \beta=\sin\frac{\vartheta_{1}}{2}\,e^{i\varphi_{1}}, \qquad \gamma=\cos\frac{\vartheta_{2}}{2}, \qquad \delta=\sin\frac{\vartheta_{2}}{2}\,e^{i\varphi_{2}},
\end{equation}
where $0\leq\vartheta_{1,2}\leq\pi$ and $0\leq\varphi_{1,2}<2\pi$. The angles $\vartheta_1$ ($\vartheta_2$) and $\varphi_1$ ($\varphi_2$) specify the unit vector ${\bm \zeta}_1$ (${\bm \zeta}_2$) of spin polarization of the ingoing (target) electron, namely
\begin{equation}
\label{spin_polarization}
{\bm\zeta}_{1(2)}=(\sin\vartheta_{1(2)}\cos\varphi_{1(2)},\sin\vartheta_{1(2)}\sin\varphi_{1(2)},\cos\vartheta_{1(2)}).
\end{equation}
Using the representation~(\ref{Bloch_sphere}), we can recast Eq.~(\ref{concurrenceresult}) into the form
\begin{eqnarray}
\label{concurrenceresult2}
C_{f}({\bm \zeta}_1,{\bm \zeta}_2)=\frac{|t_{d}||t_{e}|(1-{\bm \zeta}_1{\bm \zeta}_2)}{|t_d|^2+|t_{e}|^{2}-{\rm Re}(t_dt_e^*)
(1+{\bm \zeta}_1{\bm \zeta}_2)}. 
\end{eqnarray}
Equation~(\ref{concurrenceresult2}) can be related to the basic spin-resolved TDCSs~(\ref{TDCS_spin-resolved_basic}):
\begin{eqnarray}
\label{concurrenceresult3}
C_{f}({\bm \zeta}_1,{\bm \zeta}_2)=\frac{2\sqrt{I^{(d)}_{\uparrow\downarrow}I^{(e)}_{\uparrow\downarrow}}(1-{\bm \zeta}_1{\bm \zeta}_2)}{I_{\uparrow\downarrow}(1-{\bm \zeta}_1{\bm \zeta}_2)+I_{\uparrow\uparrow}(1+{\bm \zeta}_1{\bm \zeta}_2)}=
\frac{\sqrt{I^{(d)}_{\uparrow\downarrow}I^{(e)}_{\uparrow\downarrow}}}{{I}_{{\bm \zeta}_1,{\bm \zeta}_2}}\,(1-{\bm \zeta}_1{\bm \zeta}_2), 
\end{eqnarray}
where $I_{\uparrow\downarrow}=I^{(d)}_{\uparrow\downarrow}+I^{(e)}_{\uparrow\downarrow}$ is the spin-unresolved TDCS for $(e,2e)$ scattering with antiparallel spins, and $I_{{\bm \zeta}_1,{\bm \zeta}_2}$ is the spin-unresolved TDCS given by Eq.~(\ref{TDCS_spin-unresolved}).

If ${\bm \zeta}_1={\bm \zeta}_2$ the concurrence is $C_{f}({\bm \zeta}_1={\bm \zeta}_2)=0$, reflecting the fact that the two-electron spin state is a disentangled triplet state. On the contrary, if ${\bm \zeta}_1=-{\bm \zeta}_2$ the concurrence is
\begin{equation}
\label{concurrence_(a)}
C_{f}({\bm \zeta}_1=-{\bm \zeta}_2)=\frac{2|t_d||t_e|}{|t_d|^2+|t_e|^2}=\frac{2\sqrt{I^{(d)}_{\uparrow\downarrow}I^{(e)}_{\uparrow\downarrow}}}{I_{\uparrow\downarrow}},
\end{equation}
reaching the maximum value $C^{\rm max}_{f}=1$ when $|t_d|=|t_e|$ ($I^{(d)}_{\uparrow\downarrow}=I^{(e)}_{\uparrow\downarrow}$). The latter condition allows for formation not only of the Bell state $\Psi^-_{\rm Bell}$, as in the $t_d=t_e$ case, but also of the Bell state $\Psi^+_{\rm Bell}$ provided $t_d=-t_e$. In the intermediate case, when spin polarizations are aligned perpendicular to each other, ${\bm \zeta}_1\perp{\bm \zeta}_2$, one has for the concurrence
\begin{equation}
\label{concurrence_(a)_1}
C_{f}({\bm \zeta}_1\perp{\bm \zeta}_2)=\frac{|t_{d}||t_{e}|}{|t_d|^2+|t_{e}|^{2}-{\rm Re}(t_dt_e^*)}=
\frac{\sqrt{I^{(d)}_{\uparrow\downarrow}I^{(e)}_{\uparrow\downarrow}}}
{I},
\end{equation}
where $I=(I_{\uparrow\downarrow}+I_{\uparrow\uparrow})/2$ is the spin-averaged TDCS for $(e,2e)$ scattering with unpolarized electrons. It can be seen that $C_{f}({\bm \zeta}_1\perp{\bm \zeta}_2)\leq C_f({\bm \zeta}_1=-{\bm \zeta}_2)$, with the maximum value $C^{\rm max}_{f}=1$ realized only when $I_{\uparrow\uparrow}=0$ ($t_{d}=t_e$).

\subsection{Bell's inequality}
The phenomenon of quantum entanglement plays a fundamental role in Bell's theorem~\cite{Bell1964}, which states that any theory based on local realism is unable to reproduce all quantum mechanical predictions. In particular, quantum mechanics predicts violation of Bell's inequality for a pair of electrons in the entangled, singlet spin state, thus rejecting the principle of local realism. The violation of Bell's inequality appears to be a sufficient criterion of an entangled state, i.e., the state violating this inequality
is entangled.\footnote{At the same time, not all entangled states violate Bell's inequality.} Here we wish to inspect in this context the pair spin state~(\ref{rho_f}).

Cirel'son~\cite{Cirelson} presented an elegant formulation of Bell's inequality (see, for instance, Ref.~\cite{PopescuRoberts} for technical details). Following this formulation we consider the operator
\begin{equation}
\label{Pi}
\hat{\Pi}=\hat{A}_{1}\left(\hat{B}_{1}-\hat{B}_{2}\right)+\hat{A}_{2}\left(\hat{B}_{1}+\hat{B}_{2}\right),
\end{equation}
where the operators  $\hat{A}_{1,2}={\bf a}_{1,2}\hat{\bm\sigma}^{(1)}$ and $\hat{B}_{1,2}={\bf b}_{1,2}\hat{\bm\sigma}^{(2)}$ stand for
projections of the first $\hat{\bm\sigma}^{(1)}$ and second $\hat{\bm\sigma}^{(2)}$ electron
spin operators in detectors $A$ and $B$, respectively. The projection directions are set by the unit vectors ${\bf a}_{1}=(0,0,1)$, ${\bf a}_{2}=(1,0,0)$ and ${\bf b}_{1}=(-1/\sqrt{2},0,-1/\sqrt{2})$, ${\bf b}_{2}=(-1/\sqrt{2},0,1/\sqrt{2})$. The classical limit of Bell's inequality leads to the condition~\cite{PopescuRoberts}
\begin{equation}
\label{classical}
\langle\hat{\Pi}\rangle={\rm Tr}\left(\hat{\rho}_{f}\hat{\Pi}\right)\leq2.
\end{equation}
Violation of this inequality signals an entangled state $\hat{\rho}_{f}$. Using Eqs.~(\ref{rho_f}) and~(\ref{Pi}), we deduce that
\begin{eqnarray}
\label{Bell3}
\langle\hat{\Pi}\rangle=\sqrt{2}\,\frac{2{\rm Re}(t_dt_e^*)(1-\zeta_{1,y}\zeta_{2,y})-\left(|t_{d}|^2+|t_e|^2\right)\left({\zeta}_{1,x}{\zeta}_{2,x}+\zeta_{1,z}\zeta_{2,z}\right)}{|t_d|^2+|t_e|^2-{\rm Re}(t_dt_e^*)(1+{\bm\zeta}_1{\bm\zeta}_2)}\leq2, 
\end{eqnarray}
where we utilized the Bloch-sphere representation~(\ref{Bloch_sphere}) and expressed the Bell's inequality in terms of the components of the unit spin-polarization vectors ${\bm\zeta}_1=(\zeta_{1,x},\zeta_{1,y},\zeta_{1,z})$ and ${\bm\zeta}_2=(\zeta_{2,x},\zeta_{2,y},\zeta_{2,z})$. As anticipated, for the specific case $t_{d}=t_{e}$ the inequality is maximally violated: $\langle\hat{\Pi}\rangle=2\sqrt{2}>2$.
Equation~(\ref{Bell3}) can be presented in terms of the basic TDCSs~(\ref{TDCS_spin-resolved_basic}):
\begin{eqnarray}
\label{Bell4}
\frac{I_{\uparrow\downarrow}(1-{\bm\zeta}_1{\bm\zeta}_2)-
I_{\uparrow\uparrow}(1-\zeta_{1,y}\zeta_{2,y})}{I_{\uparrow\downarrow}(1-{\bm\zeta}_1{\bm\zeta}_2)+I_{\uparrow\uparrow}(1+{\bm\zeta}_1{\bm\zeta}_2)}\leq\frac{1}{\sqrt{2}}.
\end{eqnarray}
The advantage of this representation is that the quantities $I_{\uparrow\uparrow}$ and $I_{\uparrow\downarrow}$ are in principle measurable with polarized initial electrons without invoking spin resolution of the outgoing electrons. Moreover, it is also not necessary to carry out absolute measurements, for the knowledge of the $I_{\uparrow\uparrow}$ and $I_{\uparrow\downarrow}$ intensities on a relative scale suffices. Finally, we note that Bell's inequality~(\ref{Bell4}) depends not only on the relative orientation of the initial-electron spins but also on their orientation with respect to the detectors' axes.

%
\section{Ensemble average over electron pairs}
\label{Spin_avg}
In the $(e,2e)$ scattering experiment one measures outgoing electron pairs emerged in collisions of electrons in the incident electron beam with electrons in target systems (atoms, molecules, clusters, etc.). The result of measurements represents thus an average over the colliding electron pairs, in particular, over their impact parameters ${\bf b}$ and initial spin states~(\ref{one-electron_spin_states}). This aspect must be properly taken into account in the above formulas for the entanglement measures and Bell's inequality.

We only briefly outline the role of different impact-parameter values in the ensemble. Under typical conditions of scattering experiments, the projectile-target systems have random impact parameters. This implies a uniform impact-parameter distribution in the ensemble of the colliding pairs of the ingoing and target electrons. The average with such distribution has been already accounted for when deriving the unnormalized density matrix~(\ref{rho_f_unnorm}). Since the normalized spin density matrix $\hat{\rho}_f=\hat{\tilde\rho}_f/{\rm Tr}\hat{\tilde\rho}_f$ remained the same as in Eq.~(\ref{rho_f}), the results for the entanglement measures and Bell's inequality thus also remain unaltered. This owes to the properties of the projectile wave packet, which is assumed to be sharply peaked in momentum space as is usually realized in scattering experiments.

Let us turn to the effect of the partial spin polarization of the incident beam and the target. The ensembles of spin states of electrons in the incident beam and target systems before the $(e,2e)$ collision can be generally described by the statistical operators 
\begin{equation}
\label{statistical_operator_in}
\hat{\rho}_{1,i}=\frac{1}{2}(\hat{I}+{\bf P}_1\hat{\bm\sigma}^{(1)}), \qquad \hat{\rho}_{2,i}=\frac{1}{2}(\hat{I}+{\bf P}_2\hat{\bm\sigma}^{(2)}).
\end{equation}
Here the polarization vectors ${\bf P}_{1,2}$ ($0\leq P_{1,2}\leq1$) are the averages of the spin polarizations of the individual electrons which are in pure spin states such as those given by Eq.~(\ref{one-electron_spin_states}):
\begin{equation}
\label{one-electron_spin_states1}
|\chi_{n_1}(1)\rangle=\alpha_{n_1}|1\uparrow\rangle+\beta_{n_1}|1\downarrow\rangle, \qquad
|\eta_{n_2}(2)\rangle=\gamma_{n_2}|2\uparrow\rangle+\delta_{n_2}|2\downarrow\rangle,
\end{equation}
where $n_{1,2}$ label the electrons in the incident beam and the target systems, respectively.
Despite the fact that the one-electron spin functions preserve unitarity $|\alpha_{n_1}|^{2}+|\beta_{n_1}|^{2}=1$,
$|\gamma_{n_1}|^{2}+|\delta_{n_1}|^{2}=1$,  $\forall n_{1,2}\in N$, the values of the coefficients are different for different electrons.
Therefore, in the generic case, ensemble averaging leads to the mixed state $P_{1,2}<1$. Only in the special case when
the coefficients are equal $\alpha_{n_1}=\alpha$, $\beta_{n_1}=\beta$, $\gamma_{n_2}=\gamma$, and $\delta_{n_2}=\delta$ for $\forall n_{1,2}\in N$
the averaging procedure preserves the pure state $P_{1,2}=1$.
The ensembles of the ingoing and target electrons in the pure state case $P_{1,2}=1$ are characterized by the single spin states~(\ref{one-electron_spin_states}) which, using Eqs.~(\ref{Bloch_sphere}) and (\ref{spin_polarization}), can be presented respectively as
\begin{equation}
\label{one-electron_spin_states11}
|\chi(1)\rangle=|1\uparrow_{{\bm\zeta}_1}\rangle, \qquad |\eta(2)\rangle=|2\uparrow_{{\bm\zeta}_2}\rangle,
\end{equation}
where ${\bm\zeta}_{1,2}={\bf P}_{1,2}$ and $|\uparrow_{{\bm\zeta}}\rangle$ designates a spin-up spinor for the ${\bm\zeta}$ quantization axis, i.e., ${\bm\zeta}\hat{\bm\sigma}|\uparrow_{{\bm\zeta}}\rangle=|\uparrow_{{\bm\zeta}}\rangle$. According to Eq.~(\ref{one-electron_spin_states11}), the statistical operators~(\ref{statistical_operator_in}) for $P_{1,2}=1$ acquire the form
\begin{equation}
\label{statistical_operator_pure}
\hat{\rho}_{1,i}=|1\uparrow_{{\bm\zeta}_1}\rangle\langle1\uparrow_{{\bm\zeta}_1}|, \qquad
\hat{\rho}_{2,i}=|2\uparrow_{{\bm\zeta}_2}\rangle\langle2\uparrow_{{\bm\zeta}_2}|,
\end{equation}
explicitly showing the presence of a single state in the statistical mixtures of the ingoing and target electron spin states. 

In practice the electrons are only partially polarized ($P<1$) or even unpolarized ($P=0$). To treat this situation, we note that the statistical operators can be presented as
\begin{eqnarray}
\label{statistical_operator_mixed}
\hat{\rho}_{1,i}&=&\frac{1+P_1}{2}\,|1\uparrow_{{\bm\zeta}_1}\rangle\langle1\uparrow_{{\bm\zeta}_1}|+
\frac{1-P_1}{2}\,|1\uparrow_{-{\bm\zeta}_1}\rangle\langle1\uparrow_{-{\bm\zeta}_1}|, \nonumber\\
\hat{\rho}_{2,i}&=&\frac{1+P_2}{2}\,|2\uparrow_{{\bm\zeta}_2}\rangle\langle2\uparrow_{{\bm\zeta}_2}|+
\frac{1-P_2}{2}\,|2\uparrow_{-{\bm\zeta}_2}\rangle\langle2\uparrow_{-{\bm\zeta}_2}|,
\end{eqnarray}
where ${\bm\zeta}_{1,2}={\bf P}_{1,2}/P_{1,2}$ and we used the fact that $|\downarrow_{{\bm\zeta}}\rangle=|\uparrow_{-{\bm\zeta}}\rangle$. From Eq.~(\ref{statistical_operator_mixed}) it follows that the statistical mixture of the ingoing (target) electron spin states is composed of two orthonormal spinors: the one has spin up and the other has spin down with respect to the polarization vector ${\bf P}_{1(2)}$. The indicated spin-up and -down states are represented in the mixture with the statistical weights $w_\uparrow=(1+P_{1(2)})/2$ and $w_\downarrow=(1-P_{1(2)})/2$, respectively. This implies that we have a statistical mixture of the following four spin states of the colliding pairs before the collision:
\begin{subequations}
\label{one-electron_spin_states2}
\begin{align}
|\chi(1)\rangle&=|1\uparrow_{{\bm\zeta}_1}\rangle, & |\eta(2)\rangle&=|2\uparrow_{{\bm\zeta}_2}\rangle, &
w_{{\bm\zeta}_1,{\bm\zeta}_2}&=\frac{1}{4}\,(1+P_1)(1+P_2), \\
|\chi(1)\rangle&=|1\uparrow_{{\bm\zeta}_1}\rangle, & |\eta(2)\rangle&=|2\uparrow_{-{\bm\zeta}_2}\rangle, &
w_{{\bm\zeta}_1,{-\bm\zeta}_2}&=\frac{1}{4}\,(1+P_1)(1-P_2), \\
|\chi(1)\rangle&=|1\uparrow_{-{\bm\zeta}_1}\rangle, & |\eta(2)\rangle&=|2\uparrow_{{\bm\zeta}_2}\rangle, &
w_{-{\bm\zeta}_1,{\bm\zeta}_2}&=\frac{1}{4}\,(1-P_1)(1+P_2), \\
|\chi(1)\rangle&=|1\uparrow_{-{\bm\zeta}_1}\rangle, & |\eta(2)\rangle&=|2\uparrow_{-{\bm\zeta}_2}\rangle, &
w_{-{\bm\zeta}_1,{-\bm\zeta}_2}&=\frac{1}{4}\,(1-P_1)(1-P_2),
\end{align}
\end{subequations}
where $w_{\pm{\bm\zeta}_1,\pm{\bm\zeta}_2}$ and $w_{\pm{\bm\zeta}_1,\mp{\bm\zeta}_2}$ are the statistical weights of these states in the mixture. Accordingly, the unnormalized final density matrix is given by the statistical average
\begin{equation}
\label{rho_f_unnorm_avg}
\hat{\tilde\rho}_f({\bf P}_1,{\bf P}_2)=w_{{\bm\zeta}_1,{\bm\zeta}_2}\hat{\tilde\rho}_f({\bm\zeta}_1,{\bm\zeta}_2)+w_{{\bm\zeta}_1,-{\bm\zeta}_2}\hat{\tilde\rho}_f({\bm\zeta}_1,-{\bm\zeta}_2)+w_{{-\bm\zeta}_1,{\bm\zeta}_2}\hat{\tilde\rho}_f(-{\bm\zeta}_1,{\bm\zeta}_2)+w_{-{\bm\zeta}_1,-{\bm\zeta}_2}\hat{\tilde\rho}_f(-{\bm\zeta}_1,-{\bm\zeta}_2),
\end{equation}
where $\hat{\tilde\rho}_f(\pm{\bm\zeta}_1,\pm{\bm\zeta}_2)$ and $\hat{\tilde\rho}_f(\pm{\bm\zeta}_1,\mp{\bm\zeta}_2)$ are the unnormalized final density matrices calculated on the basis of Eqs.~(\ref{rho_f_unnorm}) and~(\ref{rho_f_unnorm_1}) using the corresponding initial pair spin states~(\ref{one-electron_spin_states2}). For the normalized density matrix we have
\begin{equation}
\label{rho_f_norm_avg}
\hat{\rho}_f({\bf P}_1,{\bf P}_2)=\frac{\hat{\tilde\rho}_f({\bf P}_1,{\bf P}_2)}{{\rm Tr}\hat{\tilde\rho}_f({\bf P}_1,{\bf P}_2)},
\end{equation}
where
\begin{equation}
\label{TDCS_avg}
{\rm Tr}\hat{\tilde\rho}_f({\bf P}_1,{\bf P}_2)=\frac{d\sigma_{{\bf P}_1,{\bf P}_2}}{dE_Bd\Omega_Ad\Omega_B}=
\frac{k_1k_2}{(2\pi)^5k_0}[|t_d|^2+|t_e|^2-(1+{\bf P}_1{\bf P}_2){\rm Re}(t_dt_e^*)]
\end{equation}
is the spin-unresolved TDCS for the initial spin polarizations ${\bf P}_{1,2}$. The explicit form of the density matrix~(\ref{rho_f_norm_avg}) is presented in Appendix~\ref{density_matrix_2-el}. In contrast to Eq.~(\ref{rho_f}), it describes a mixed pair state, which reduces to the pure state only in specific cases, for instance, when $t_d=t_e$ (or, equivalently, $I_{\uparrow\uparrow}=0$). As opposed to the result~(\ref{concurrenceresult2}) obtained for the pure pair state, a general expression for the pair concurrence defined in Eq.~(\ref{conurrence_def}) becomes too cumbersome. For this reason, here we restrict ourselves with the cases where one electron is unpolarized, $P_{1(2)}=1$ with $P_{2(1)}=0$, and where both electrons are unpolzarized, ${P}_{1,2}=0$, which is most studied in $(e,2e)$ scattering experiments. For the first case we derive the pair concurrence the same as in Eq.~(\ref{concurrence_(a)_1}), and for the second case we get
\begin{equation}
\label{C_out_unpolarized}
C_{f}(P_{1,2}=0)=\theta\left(|t_d+t_e|^2-3|t_d-t_e|^2\right)\,\frac{4{\rm Re}(t_dt_e^*)-|t_d|^2-|t_e|^2}{2[|t_d|^2+|t_e|^2-{\rm Re}(t_dt_e^*)]},
\end{equation}
where $\theta$ is the Heaviside step function.
Introducing the singlet-channel $I_{\rm s}$ and triplet-channel $I_{\rm t}$ components of the spin-averaged TDCS with unpolarized electrons according to the relations
$$
\frac{d\sigma_{{\bf P}_{1,2}=0}}{dE_2d\Omega_1\Omega_2}=I_{\rm s}+I_{\rm t}, \qquad
I_{\rm s}=\frac{1}{4}\,\frac{k_1k_2}{(2\pi)^5k_0}\,|t_d+t_e|^2, \qquad
I_{\rm t}=\frac{3}{4}\,I_{\uparrow\uparrow}=\frac{3}{4}\,\frac{k_1k_2}{(2\pi)^5k_0}\,|t_d-t_e|^2,
$$
we may express Eq.~(\ref{C_out_unpolarized}) in the form
\begin{equation}
\label{C_out_unpolarized_1}
C_f(P_{1,2}=0)=\theta(I_{\rm s}-I_{\rm t})\,\frac{I_{\rm s}-I_{\rm t}}{I_{\rm s}+I_{\rm t}}.
\end{equation}
This result shows that the pair concurrence for unpolarized initial electrons turns out to be nonzero only when the singlet-channel scattering dominates, namely $I_{\rm s}>I_{\rm t}$. The maximum value $C_f^{\rm max}=1$ is reached when the triplet scattering is absent, i.e., $I_{\rm t}=0$, that amounts to $t_d=t_e$.

Let us consider Bell's inequality. Using the density matrix~(\ref{rho_f_norm_avg}) in Eq.~(\ref{classical}), we derive
\begin{eqnarray}
\label{Bell5}
\frac{I_{\uparrow\downarrow}(1-{\bf P}_1{\bf P}_2)-
I_{\uparrow\uparrow}(1-P_{1,y}P_{2,y})}{I_{\uparrow\downarrow}(1-{\bf P}_1{\bf P}_2)+I_{\uparrow\uparrow}(1+{\bf P}_1{\bf P}_2)}\leq\frac{1}{\sqrt{2}}.
\end{eqnarray}
It is identical to Eq.~(\ref{Bell4}), except for the replacement of the unit vectors ${\bm\zeta}_{1,2}$ 
with the polarization vectors ${\bf P}_{1,2}$. When at least one of the electrons is unpolarized ($P_{1}=0$ and/or $P_2=0$) the inequality takes the form
\begin{eqnarray}
\label{Bell6}
\mathcal{A}\leq\frac{1}{\sqrt{2}}, \qquad \mathcal{A}=\frac{I_{\uparrow\downarrow}-I_{\uparrow\uparrow}}{I_{\uparrow\downarrow}+I_{\uparrow\uparrow}},
\end{eqnarray}
where $\mathcal{A}$ is the so-called spin asymmetry for $(e,2e)$ scattering. Measuring $\mathcal{A}>1/\sqrt{2}$ thus provides an indication of the violation of Bell's inequality.

\section{Scattering amplitudes}
\label{scat_amplitude}
For calculating both the entanglement measures and the TDCSs for the $(e,2e)$ collision we need the knowledge of the direct and exchange scattering amplitudes $t_d$ and $t_e$. According to the above discussion, the maximal effect of entanglement of an outgoing electron pair appears to be always realized if $t_d=t_e\neq0$. Therefore, it is interesting to determine such situations where one has for the amplitudes $t_d=t_e$, even not knowing their exact values. For this purpose one can make use of the various symmetry transformations under which the projectile-target system remains invariant, for example, such as translation, rotation, inversion, time reversal, etc. We focus on the parity transformation which is equivalent to the combined inversion and rotation operations. In the discussed case of a nonrelativistic projectile-target system without spin-orbit couplings this transformation leaves unchanged the interaction potentials between the fragments both in the initial and in the final channel. Consider the reflection of electron coordinates with respect to the plane containing the incident electron momentum ${\bf k}_0$. If upon the indicated reflection the outgoing electron momenta transform as ${\bf k}_{A,B}\to{\bf k}'_{A,B}$, then the direct and exchange amplitudes~(\ref{direct&exchange}) are
\begin{equation}
\label{direct&exchange_reflect}
t_d=t({\bf k}_0,\psi_{\mathcal T}'\to{\bf k}_A',{\bf k}_B'), \qquad t_e=t({\bf k}_0,\psi_{\mathcal T}'\to{\bf k}_B',{\bf k}_A'),
\end{equation}
where $\psi_{\mathcal T}'=\hat{\mathcal{P}}\psi_{\mathcal T}$ is the target wave function after the action of the corresponding parity operator $\hat{\mathcal{P}}$. In the case of symmetric kinematics (equal energy sharing $E_A=E_B$ and ${\bf k}_A{\bf k}_0={\bf k}_B{\bf k}_0$) we can always choose a mirror plane such that ${\bf k}_A'={\bf k}_B$ and ${\bf k}_B'={\bf k}_A$. Further, if the target Hamiltonian $\hat{H}_{\mathcal T}$ is invariant under reflection with respect to this plane, i.e., $[\hat{H}_{\mathcal T},\hat{\mathcal{P}}]=0$, then $\psi_{\rm A}'={\mathcal{P}_{\mathcal T}}\psi_{\mathcal T}$, with $\mathcal{P}_{\mathcal T}=+1$ (even parity) or $\mathcal{P}_{\mathcal T}=-1$ (odd parity). Hence, in Eq.~(\ref{direct&exchange_reflect}), we have
\begin{equation}
\label{direct&exchange_reflect1}
t_d=\mathcal{P}_{\mathcal T}t({\bf k}_0,\psi_{\mathcal T}\to{\bf k}_B,{\bf k}_A), \qquad
t_e=\mathcal{P}_{\mathcal T}t({\bf k}_0,\psi_{\mathcal T}\to{\bf k}_A,{\bf k}_B).
\end{equation}
This leads to the relations $t_d-t_e=-\mathcal{P}_{\mathcal T}(t_d-t_e)$ and $t_d+t_e=\mathcal{P}_{\mathcal T}(t_d+t_e)$, so that either $t_d=t_e$ or $t_d=-t_e$ depending on whether the parity $\mathcal{P}_{\mathcal T}$ of the target wave function $\psi_{\mathcal T}$ is even or odd, respectively. In the remainder of this section we outline some specific approximations for the $t_{d,e}$ amplitudes.

\subsection{$(e,2e)$ electron momentum spectroscopy}
\label{EMS}
Calculation of $t_d$ and $t_e$ strongly simplifies in a particular case of the $(e,2e)$ process which is usually referred to as electron momentum spectroscopy (EMS)~\cite{Weigold_book}. The marked feature of EMS is the $(e,2e)$ kinematics close to the kinematical regime of a free electron-electron collision. This validates the plane-wave Born approximation for the $T$ matrix. The direct and exchange amplitudes~(\ref{direct&exchange}) in this approximation acquire the forms
\begin{equation}
\label{direct&exchange_PWBA}
t_d=\frac{4\pi}{|{\bf k}_0-{\bf k}_A|^2}\,\phi_{\mathcal T}({\bf q}), \qquad t_e=\frac{4\pi}{|{\bf k}_0-{\bf k}_B|^2}\,\phi_{\mathcal T}({\bf q}),
\end{equation}
where $\phi_{\mathcal T}({\bf q})=\langle{\bf q}|\psi_{\mathcal T}\rangle$ is the momentum-space wave function of the target electron, and ${\bf q}={\bf k}_A+{\bf k}_B-{\bf k}_0$. It follows that in the kinematical regime $|{\bf k}_0-{\bf k}_A|=|{\bf k}_0-{\bf k}_B|$ or, equivalently, $|{\bf k}_A-{\bf q}|=|{\bf k}_B-{\bf q}|$, one has $t_d=t_e$. At the same time, it should be noted that in this kinematical regime one can encounter $t_d=t_e=0$ at equal energy sharing $E_A=E_B$ if the wave function $\phi_{\mathcal T}({\bf q})$ has odd parity ($\mathcal{P}_{\mathcal T}=-1$) in Eq.~(\ref{direct&exchange_reflect1}).

\subsection{$(e,2e)$ on atomic hydrogen in the 3C model}
Atomic hydrogen is a benchmark target for $(e,2e)$ ionization: (i) the hydrogen state $|\phi_{{\rm H}(1s)}\rangle$ is exactly known and (ii) the scattering problem is a three-body Coulomb problem. The $T$ matrix in this case is given by
\begin{equation}
\label{T-matrix_H}
t({\bf k}_0,\psi_{{\rm H}(1s)}\to{\bf k}_A,{\bf k}_B)=\langle\Psi^{(-)}_{{\bf k}_A,{\bf k}_B}|\frac{1}{r_{12}}-\frac{1}{r_1}|{\bf k}_0,\psi_{{\rm H}(1s)}\rangle,
\end{equation}
where $r_{12}=|{\bf r}_1-{\bf r}_2|$ and $r_1$ are the distances between the ingoing electron and the atomic electron and nucleus (proton). $|\Psi^{(-)}_{{\bf k}_A,{\bf k}_B}\rangle$ is a (time-reversed) scattering state of the three-body system composed of two electrons and a proton. It is a solution of the Schr\"odinger equation with the proper three-body Coulomb asymptotic condition. Finding such a solution is, in general, an intractable task. Therefore, we resort to a well-known model usually referred to as 3C (or BBK)~\cite{BBK1989}. In the 3C model, one employs the three-body scattering state in the form
\begin{equation}
\label{3C}
\Psi^{(-)}_{{\bf k}_A,{\bf k}_B}({\bf r}_1,{\bf r}_2)=\psi^{(-)}_{{\bf k}_A}({\bf r}_1;Z=1)\psi^{(-)}_{{\bf k}_B}({\bf r}_2;Z=1)
f^{(-)}_{{\bf k}_{AB}}({\bf r}_{12}),
\end{equation}
where 
\begin{equation}
\label{Coulomb_wave}
\psi^{(-)}_{{\bf k}}({\bf r};Z)=e^{-\pi\xi/2}\Gamma(1-i\xi)e^{i{\bf k}{\bf r}}{_1}F_1(i\xi,1;-ikr-i{\bf k}{\bf r})
\end{equation}
is the Coulomb wave, with $\xi=-Z/k$, that describes the scattering state in the electron-nucleus pair, and the electron-electron Coulomb correlation factor is
\begin{equation}
\label{ee_corr_factor}
f^{(-)}_{{\bf k}_{AB}}({\bf r}_{12})=e^{-\pi\xi_{AB}/2}\Gamma(1-i\xi_{AB}){_1}F_1(i\xi_{AB},1;-i{k}_{AB}r_{12}-i{\bf k}_{AB}{\bf r}_{12}),
\end{equation}
where ${\bf k}_{AB}=({\bf k}_A-{\bf k}_B)/2$ and $\xi_{AB}=1/2k_{AB}$. Using the 3C function~(\ref{3C}) in Eq.~(\ref{T-matrix_H}), we get the following six-dimensional integral for the $T$ matrix:
\begin{eqnarray}
\label{T-matrix_H_3C}
t({\bf k}_0,\psi_{{\rm H}(1s)}\to{\bf k}_A,{\bf k}_B)&=&\int d{\bf r}_1\int d{\bf r}_2\,\psi^{(-)*}_{{\bf k}_A}({\bf r}_1;Z=1)
\psi^{(-)*}_{{\bf k}_B}({\bf r}_2;Z=1)
\nonumber\\
&{}&\times f^{(-)*}_{{\bf k}_{AB}}({\bf r}_{12})\left(\frac{1}{r_{12}}-\frac{1}{r_1}\right)e^{i{\bf k}_0{\bf r}_1}\psi_{1s}({\bf r}_2;Z=1),
\end{eqnarray}
with the $1s$ wave function
\begin{equation}
\label{1s_wf}
\psi_{1s}({\bf r}_2;Z)=\sqrt{\frac{Z^3}{\pi}}\,e^{-Zr_2}.
\end{equation}
We use the 3C model~(\ref{T-matrix_H_3C}) in the next section devoted to numerical results for $(e,2e)$ ionization of atomic hydrogen. In this regard it should be noted that the integration in Eq.~(\ref{T-matrix_H_3C}) can be reduced to a two-dimensional one (see, for instance, Appendix 2 of Ref.~\cite{BBK1989}). Since the $1s$ hydrogen state has even parity, in symmetric kinematics we should have $t_d=t_e$. It can be easily verified that the 3C model~(\ref{T-matrix_H_3C}) obeys this condition.

\section{Results and discussion}
\label{res}
For illustration purposes, we consider the $(e,2e)$ ionization of atomic hydrogen
$$
e^-+{\rm H}\to{\rm H}^++2e^-
$$
in the coplanar kinematics with equal energy sharing that favors the formation of spin entanglement. The incident electron energy is chosen to be $E_0=54.4$ eV (2 a.u.) so that the ionization yield is close to maximum~\cite{Erhardt1986}. Accordingly, the final electron energies are $E_1=E_2=20.4$ eV (0.75 a.u.). Using the 3C model~(\ref{T-matrix_H_3C}), which is more valid for the chosen incident energy than the plane-wave Born approximation~(\ref{direct&exchange_PWBA}), we calculate numerically the TDCS, pair concurrence, and entanglement of formation as functions of the electron in-plane angles $\theta_A$ and $\theta_B$ measured with respect to the direction of the incident electron momentum ${\bf k}_0$. These angles vary from $-180^\circ$ to $180^\circ$, so that the angular ranges
$0^\circ\leq\theta_{A,B}\leq180^\circ$ and $-180^\circ\leq\theta_{A,B}\leq0^\circ$ correspond to upper and lower half-planes, respectively. In addition, we examine the violation of Bell's inequality in the scattering plane numerically. It should be noted that the criteria of entanglement can signal its maximal effect even when the TDCS value is rather small and falls beyond the experimental sensitivity. For mimicking such a circumstance, we set a threshold of $0.05\times$max(TDCS): below this threshold, all the quantities become ``unmeasurable'' or, from the viewpoint of an experiment, have zero values.

We wish to inspect different cases of electron spin polarizations in the initial channel. Here we study the following four different situations in terms of the spin polarization vectors ${\bf P}_{1,2}$ of the ingoing and target electrons: (i) $P_{1,2}=1$ with ${\bf P}_1\perp{\bf P}_2$, (ii) $P_{1,2}=1$ with ${\bf P}_1=-{\bf P}_2$, (iii) $P_{1(2)}=1$ with $P_{2(1)}=0$, and (iv) $P_{1,2}=0$. Note that we do not consider the trivial case $P_{1,2}=1$ with ${\bf P}_1={\bf P}_2$, where both the pair concurrence and the entanglement of formation are zero irrespective of kinematical conditions.

\subsection{The pair concurrence and entanglement of formation}
\begin{figure}
\includegraphics[width=\textwidth]{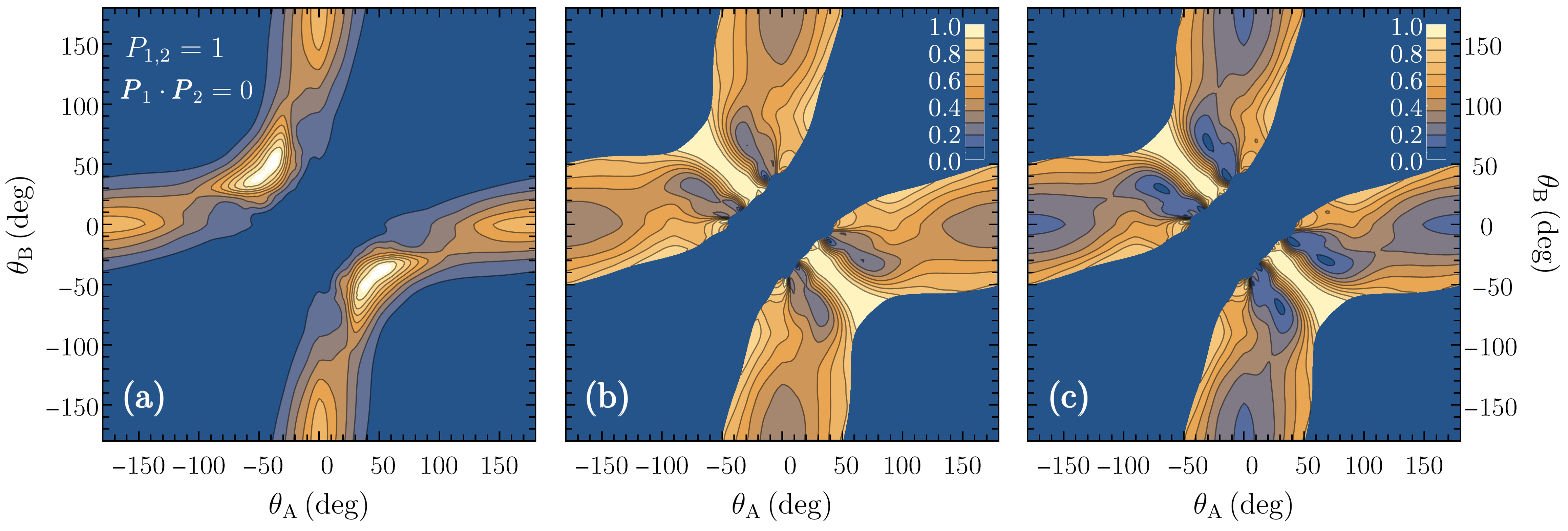}
\caption{\label{fig1} Spin-unresolved TDCS (a), pair concurrence (b), and entanglement of formation (c) as functions of the electron emission angles $\theta_{A,B}$ in the scattering plane when $P_{1,2}=1$ with ${\bf P}_1\perp{\bf P}_2$. The nonzero values are shown only in the areas where TDCS$\geq0.05\times$max(TDCS).}
\end{figure}

\begin{figure}
\includegraphics[width=\textwidth]{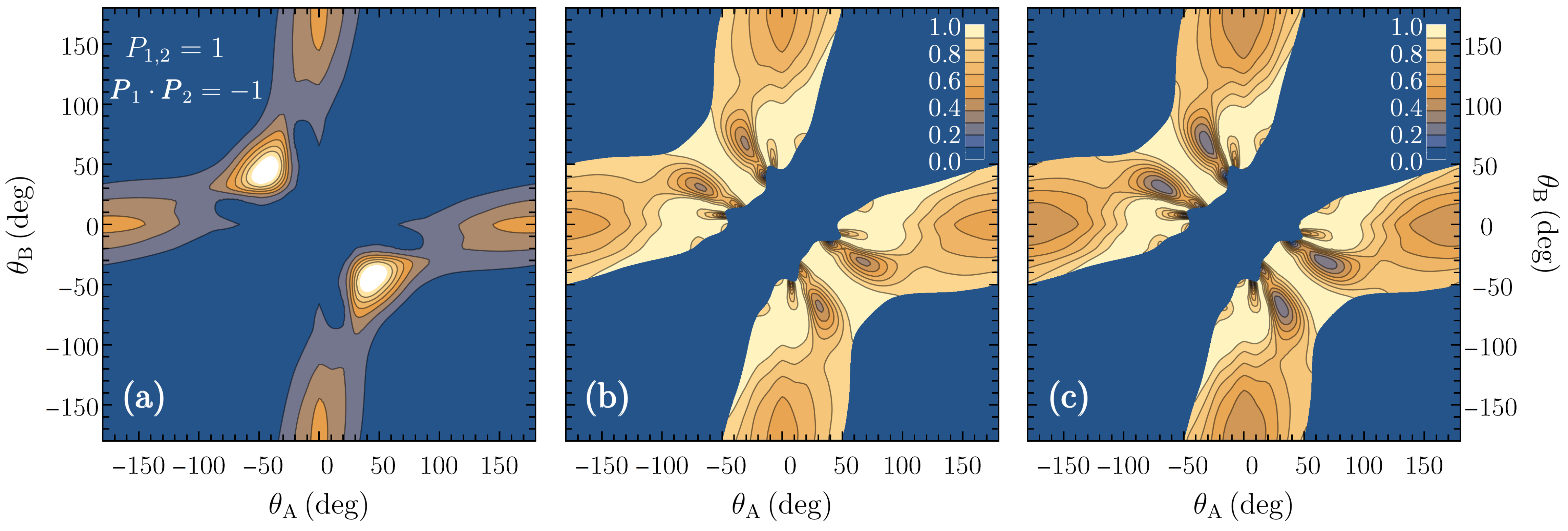}
\caption{\label{fig2} Same as in Fig.~\ref{fig1}, but for ${\bf P}_1=-{\bf P}_2$.}
\end{figure}

The numerical results when both electrons are fully polarized ($P_{1,2}=1$) are presented in Figs.~\ref{fig1} and~\ref{fig2}. As anticipated, they exhibit symmetric patterns with respect to the $\theta_A=\pm\theta_B$ lines. The TDCS appears to be peaked in the regime $\theta_A=-\theta_B\sim\pm\pi/4$, which is kinematically close to a free electron-electron collision at equal energy sharing, where owing to the energy and momentum conservation laws one has $E_A=E_B=E_0/2$ and $\theta_A=-\theta_B=\pm\pi/4$. Due to the Coulomb repulsion the electrons are preferably emitted in different half-planes. Some structures also can be seen in the TDCS when the one outgoing electron is emitted in the vicinity of the forward (backward) direction, while the other in the vicinity of the backward (forward) direction.
The pair concurrence and entanglement of formation behave similar to each other, for the latter is the convex function~(\ref{Entanglement_formation}) of the former with $E_F(0)=0$ and $E_F(1)=1$. They take the maximal value of unity if $\theta_A=-\theta_B$, where one has $t_d=t_e$. The role of the mutual orientation of initial electron-spin polarizations in entanglement of the outgoing electron pair can be seen from a comparison of Figs.~\ref{fig1} and~\ref{fig2}: for the antiparallel orientation ${\bf P}_1=-{\bf P}_2$ the effect of entanglement is much stronger manifested than for the perpendicular orientation ${\bf P}_1\perp{\bf P}_2$. This observation is readily explained by maximization of the relative contribution of the singlet state $\Psi^{-}_{\rm Bell}$ to the final pair state if ${\bf P}_1=-{\bf P}_2$.

\begin{figure}
\includegraphics[width=\textwidth]{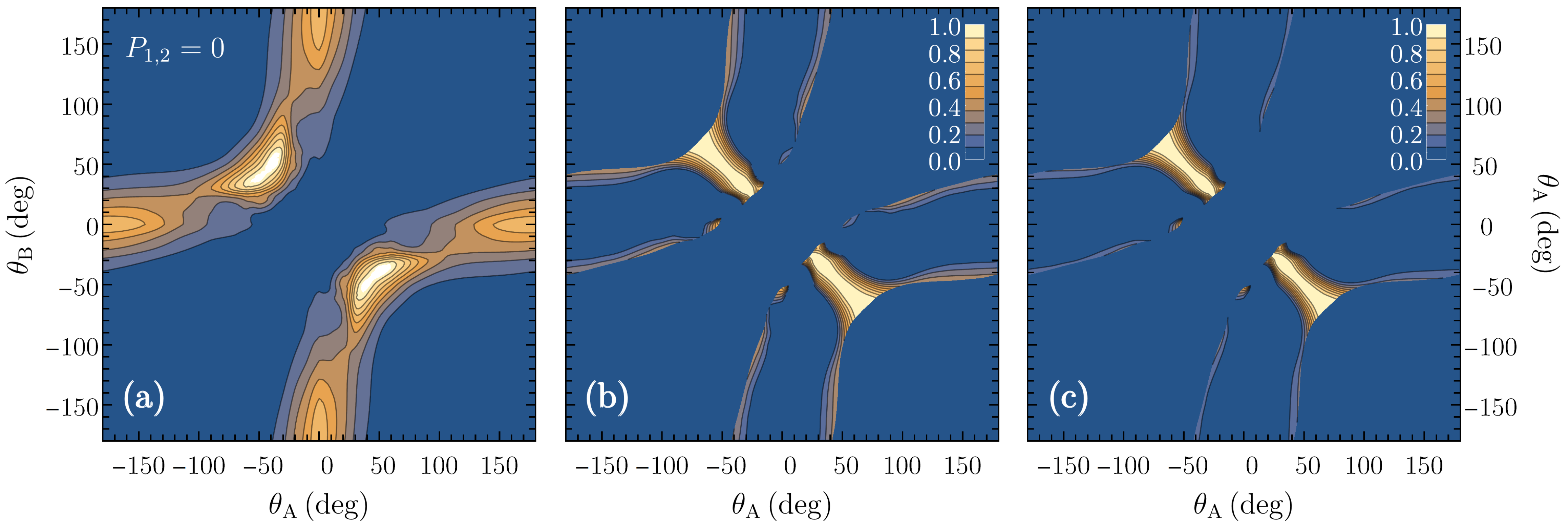}
\caption{\label{fig3} Same as in Fig.~\ref{fig1}, but for ${P}_{1,2}=0$.}
\end{figure}

The TDCS, pair concurrence, and entanglement of formation for the case of one unpolarized electron ($P_{1(2)}=1$ with $P_{2(1)}=0$) are the same as in Fig.~\ref{fig1} and therefore are not presented here. Figure~\ref{fig3} shows the results when both electrons are unpolarized. The TDCS is the same as in the case of $P_{1,2}=1$ with ${\bf P}_1\perp{\bf P}_2$ shown in Fig.~\ref{fig1}(a). However, the entanglement measures behave markedly different. Practically everywhere in the scattering plane, the pair concurrence and entanglement of formation are zero, except for those regions in the vicinity of the $\theta_A=-\theta_B$ line where the TDCS maximum is observed. In this regard, it should be noted that a common feature of the results presented in Figs.~\ref{fig1}--\ref{fig3} lies in the overlap of the maximum values of TDCS with those of the pair concurrence and entanglement of formation. This can serve as a hint for experimental investigation of the entanglement effects in the discussed $(e,2e)$ scattering process.
\subsection{Bell's inequality violation}
\begin{figure}
\includegraphics[width=\textwidth]{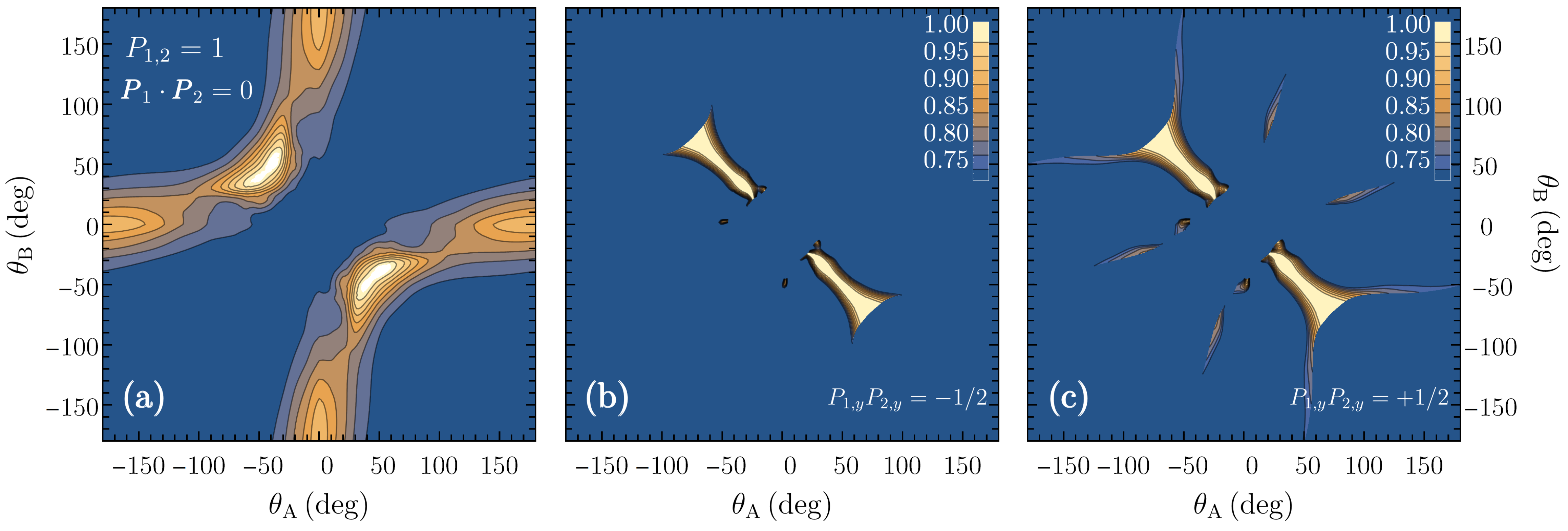}
\caption{\label{fig4} Spin-unresolved TDCS (a) and the left-hand side of Bell's inequality~(\ref{Bell5}) [panels (b) and (c)] as functions of the electron emission angles $\theta_{A,B}$ in the scattering plane when $P_{1,2}=1$ with ${\bf P}_1\perp{\bf P}_2$. The nonzero values are shown only in the angular regions where TDCS$\geq0.05\times$max(TDCS).}
\end{figure}

\begin{figure}
\includegraphics[width=\textwidth]{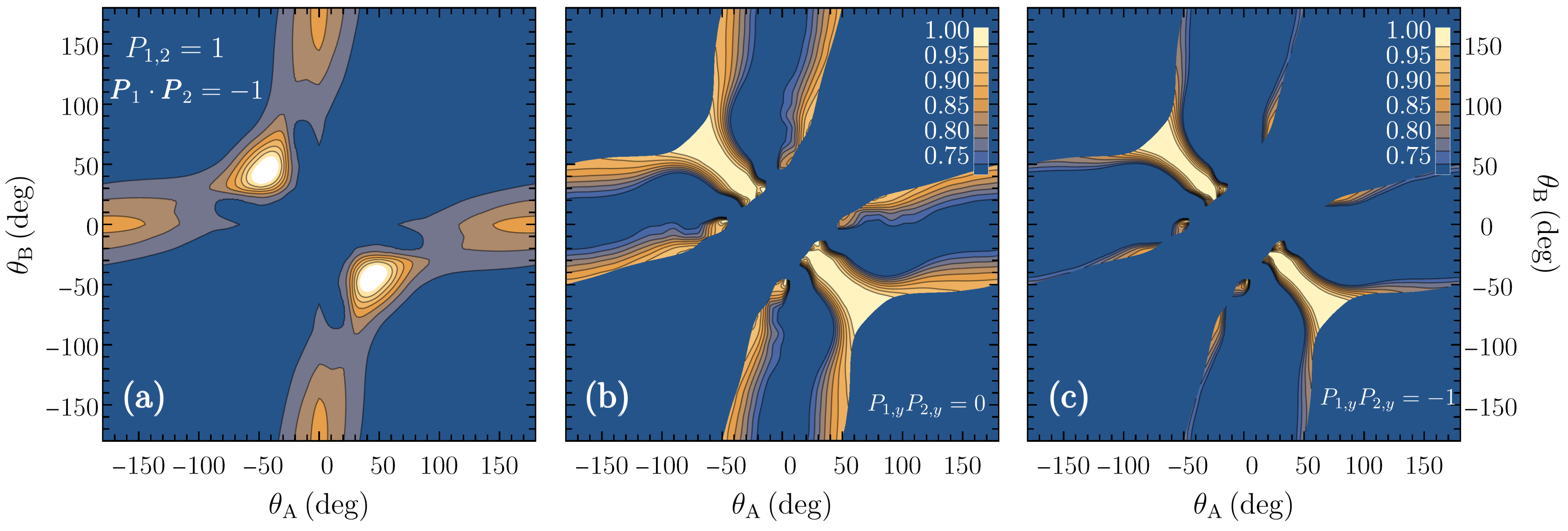}
\caption{\label{fig5} Same as in Fig.~\ref{fig4}, but for ${\bf P}_1=-{\bf P}_2$.}
\end{figure}

\begin{figure}
\includegraphics[width=\textwidth]{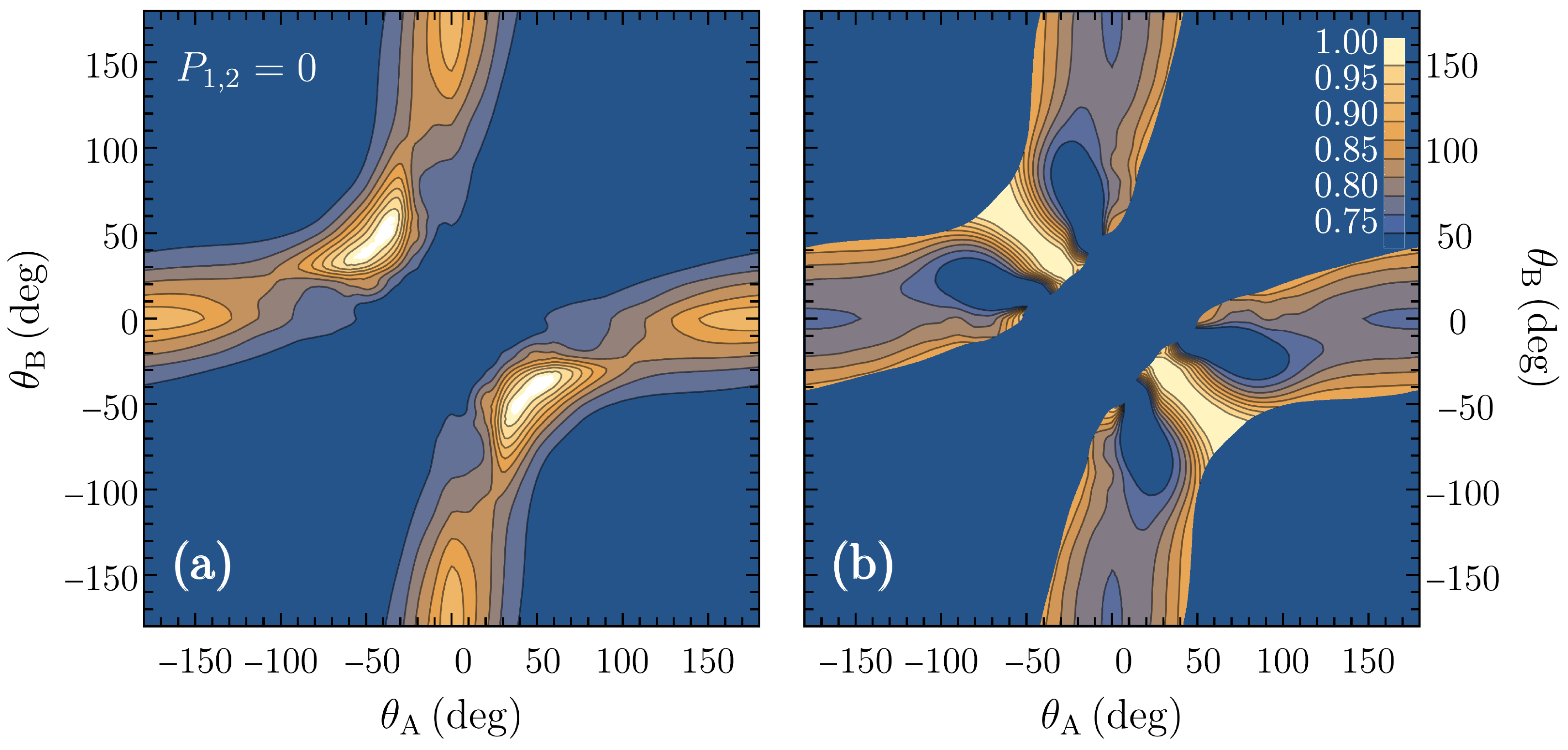}
\caption{\label{fig6} Spin-unresolved TDCS (a) and the spin asymmetry $\mathcal{A}$ (b) as functions of the electron emission angles $\theta_{A,B}$ in the scattering plane for $P_{1,2}=0$. The results for the case of $P_{1(2)}=1$ with $P_{2(1)}=0$ are identical. The light areas in panel (b) show the angular regions where the spin asymmetry violates Bell's inequality $\mathcal{A}\leq1/\sqrt{2}$ and is ``measurable.''}
\end{figure}

Experimental tests of quantum entanglement of particle pairs usually are based on measuring the violation of Bell's inequality. Figures~\ref{fig4} and~\ref{fig5} present numerical results for the Bell's inequality in the scattering plane of the $(e,2e)$ collision on atomic hydrogen in the case of polarized incident and target electrons ($P_{1,2}=1$).
The light areas in panels (b) and (c) represent the angular domains in the scattering plane where Bell's inequality is both ``measurable'' and violated. Since among the Bell states only the singlet state $\Psi^-_{\rm Bell}$ violates the inequality, the violation is more pronounced for ${\bf P}_1=-{\bf P}_2$. The violation can also be controlled by tuning the $P_{1,y}P_{2,y}$ value. Such controlling is absent when at least one of the two electrons in the initial channel of the $(e,2e)$ collision is unpolarized. Figure~\ref{fig6} shows the results in this case, taking into account that the left-hand side of Bell's inequality~(\ref{Bell5}) is equivalent to the spin asymmetry~(\ref{Bell6}). It can be seen that the total area of angular regions violating the inequality turns out to be even larger than in Fig.~\ref{fig5}(b). This is due to the difference in the ``measurable'' TDCS. Finally, similar to Figs.~\ref{fig1}--\ref{fig3}, where one has overlapping of the maximal values of the TDCS with those of the pair concurrence and entanglement of formation, the TDCS maxima in Figs.~\ref{fig4}--\ref{fig6} overlap with the maximal violation of Bell's inequality.
\section{Summary and concluding remarks}
\label{concl}
In this work, we have developed a theoretical apparatus for treating the quantum electron-pair entanglement in the nonrelativistic $(e,2e)$ collisions. The framework of the time-dependent scattering theory has been employed to elucidate how the spin entanglement of the electron pair can emerge as a result of the $(e,2e)$ scattering process. This also helped us avoiding the confusion about the entanglement in the initial channel of the process that arises in the time-independent formalism due to the delocalized incident electron states. We have derived the final spin state of the electron pair and quantified its entanglement with the pair concurrence and entanglement of formation. We have obtained the analytical expressions for these entanglement measures in terms of both the $(e,2e)$ scattering amplitudes and the $(e,2e)$ scattering cross sections with polarized electrons. We also have expressed Bell's inequality in the indicated terms. The problem of averaging over impact parameters and spin states of the projectile-target systems has been addressed and the ensemble average for the entanglement measures and Bell's inequality has been carried out using the formalism of the electron-pair spin density matrix. We have outlined symmetry properties of the direct and exchange $(e,2e)$ scattering amplitudes and formulated the well-known approximations for their evaluation such as the plane-wave Born approximation and 3C model. Using the 3C model, we have performed numerical calculations of the pair concurrence and entanglement of formation for the outgoing electron pair in the $(e,2e)$ ionization of atomic hydrogen at equal energy sharing. At that, various spin polarizations of the ingoing and target electrons have been inspected. The violation of Bell's inequality has been also investigated numerically. It has been found that the areas of the scattering plane where the TDCS is peaked overlap with those where both the entanglement measures and the violation of Bell's inequality are maximal.

The above observation can be useful for experimental tests of the entanglement phenomena in the discussed ionization process, in particular for measuring the violation of Bell's inequality. Currently, such studies are beyond the capabilities of the state-of-the-art $(e,2e)$ spectroscopy technique. Nevertheless, the analytical expressions for the entanglement measures and Bell's inequality obtained in this work show that the entanglement of the outgoing electron pair can be already quantified in $(e,2e)$ experiments by studying the basic spin-resolved TDCSs. It is important that such quantification is free from the theoretical uncertainties associated with the approximations involved in the calculations of the scattering amplitudes. Finally, it is also worth noting that the present theoretical formulation is not limited to the applications in $(e,2e)$ spectroscopy and, in principle, can be extended to a more general case, for example, to the studies of electron-electron collisions in solids and plasmas.

\begin{acknowledgments}
We are grateful to Yuri Popov, Sergey Samarin, Alexei Grum-Grzhimailo, and Nikolai Nikitin for useful discussions. K. A. Kouzakov expresses his sincere gratitude for hospitality during his research stay at Max-Planck Institute for Microstructure Physics (Halle, Germany) in August of 2018.
\end{acknowledgments}
\appendix
\section{Integrals with the ingoing wave packet}
\label{integral_wave_packet}
We perform integrations in Eq.~(\ref{rho_f_unnorm}) following the approach to the derivation of the cross section for potential scattering of a wave packet formulated in the textbook of Taylor~\cite{Taylor_book}. First, we consider the integral in the plane of the impact parameters ${\bf b}$:
\begin{eqnarray}
\label{over_b}
J=\int d^2b\,|\mathcal{F}({\bf k}_0,{\bf b};E_A,E_B,E_{\mathcal T})|^2&=&\int d^2b\int\frac{d^3 p}{(2\pi)^2}\int\frac{d^3 p'}{(2\pi)^2}\,e^{-i({\bf p}-{\bf p}'){\bf b}}\phi_{{\bf k}_0}({\bf p})\phi^*_{{\bf k}_0}({\bf p}')\nonumber\\
&{}&\times\delta\left(\frac{p^2}{2}+E_{\mathcal T}-E_A-E_B\right)\delta\left(\frac{p'^2}{2}+E_{\mathcal T}-E_A-E_B\right).\nonumber\\
\end{eqnarray}
Using
$$
\int d^2b\,e^{-i({\bf p}-{\bf p}'){\bf b}}=(2\pi)^2\delta^{(2)}({\bf p}_\perp-{\bf p}'_\perp)
$$
and the relation $\delta(a-b)\delta(b-c)=\delta(a-c)\delta(b-c)$, we obtain
\begin{eqnarray}
\label{over_b_1}
J=\int\frac{d^3 p}{(2\pi)^2}\,\delta\left(\frac{p^2}{2}+E_{\mathcal T}-E_A-E_B\right)\int\limits_{-\infty}^\infty dp'_\parallel\,\phi_{{\bf k}_0}({\bf p}_\perp,p_\parallel)\phi^*_{{\bf k}_0}({\bf p}_\perp,p'_\parallel)
\delta\left(\frac{p^2_\parallel}{2}-\frac{p'^2_\parallel}{2}\right).
\end{eqnarray}
We have
$$
\delta\left(\frac{p^2_\parallel}{2}-\frac{p'^2_\parallel}{2}\right)=\frac{1}{|p_\parallel|}
[\delta(p_\parallel-p_\parallel')+\delta(p_\parallel+p_\parallel')]=\frac{1}{k_0}\,\delta(p_\parallel-p_\parallel'),
$$
where the latter equality owes to a sharply peaked about ${\bf k}_0$ wave packet. This leads to
\begin{eqnarray}
\label{over_b_2}
J=\frac{1}{k_0}\int\frac{d^3 p}{(2\pi)^2}\,\delta\left(\frac{p^2}{2}+E_{\mathcal T}-E_A-E_B\right)\,|\phi_{{\bf k}_0}({\bf p})|^2. 
\end{eqnarray}
Finally, the integration over $E_A$ in Eq.~(\ref{rho_f_unnorm}) removes the delta function in the above integral, so that we are left with the normalization integral
$$
\int\frac{d^3 p}{(2\pi)^3}\,|\phi_{{\bf k}_0}({\bf p})|^2=1.
$$
\section{Two-electron density matrices}
\label{density_matrix_2-el}
We present the final pair density matrices~(\ref{rho_f}) and~(\ref{rho_f_norm_avg}) using the basis of Bell's states:
\begin{equation}
\label{Bell's_basis}
|\Phi^+_{\rm Bell}\rangle=
\begin{pmatrix}
1\\0\\0\\0
\end{pmatrix}, \qquad
|\Phi^-_{\rm Bell}\rangle=
\begin{pmatrix}
0\\1\\0\\0
\end{pmatrix}, \qquad
|\Psi^+_{\rm Bell}\rangle=
\begin{pmatrix}
0\\0\\1\\0
\end{pmatrix}, \qquad
|\Psi^-_{\rm Bell}\rangle=
\begin{pmatrix}
0\\0\\0\\1
\end{pmatrix}.
\end{equation}
Employing the Bloch-sphere representation~(\ref{Bloch_sphere}) and (\ref{spin_polarization}), from Eq.~(\ref{rho_f}) we derive
\begin{eqnarray}
\label{out_spin_density_matrix_pure}
(\rho_{f})_{11}&=&\frac{1}{4u}\,|t_d-t_e|^2
(1+\zeta_{1,x}\zeta_{2,x}-\zeta_{1,y}\zeta_{2,y}+\zeta_{1,z}\zeta_{2,z}), \nonumber\\
(\rho_{f})_{12}&=&(\rho_{f}^*)_{21}=\frac{1}{4u}\,|t_d-t_e|^2
(\zeta_{1,z}+\zeta_{2,z}+i\zeta_{1,x}\zeta_{2,y}+i\zeta_{1,y}\zeta_{2,x}), \nonumber\\
(\rho_{f})_{13}&=&(\rho_{f}^*)_{31}=\frac{1}{4u}\,|t_d-t_e|^2
(\zeta_{1,x}+\zeta_{2,x}-i\zeta_{1,y}\zeta_{2,z}-i\zeta_{1,z}\zeta_{2,y}), \nonumber\\
(\rho_{f})_{14}&=&(\rho_{f}^*)_{41}=\frac{1}{4u}\,(t_d-t_e)(t_d^*+t_e^*)
(i\zeta_{1,y}-i\zeta_{2,y}-\zeta_{1,x}\zeta_{2,z}+\zeta_{1,z}\zeta_{2,x}), \nonumber\\
(\rho_{f})_{22}&=&\frac{1}{4u}\,|t_d-t_e|^2
(1-\zeta_{1,x}\zeta_{2,x}+\zeta_{1,y}\zeta_{2,y}+\zeta_{1,z}\zeta_{2,z}), \nonumber\\
(\rho_{f})_{23}&=&(\rho_{f}^*)_{32}=\frac{1}{4u}\,|t_d-t_e|^2
(-i\zeta_{1,y}-i\zeta_{2,y}+\zeta_{1,x}\zeta_{2,z}+\zeta_{1,z}\zeta_{2,x}), \nonumber\\
(\rho_{f})_{24}&=&(\rho_{f}^*)_{42}=\frac{1}{4u}\,(t_d-t_e)(t_d^*+t_e^*)
(-\zeta_{1,x}+\zeta_{2,x}+i\zeta_{1,y}\zeta_{2,z}-i\zeta_{1,z}\zeta_{2,y}), \nonumber\\
(\rho_{f})_{33}&=&\frac{1}{4u}\,|t_d-t_e|^2
(1+\zeta_{1,x}\zeta_{2,x}+\zeta_{1,y}\zeta_{2,y}-\zeta_{1,z}\zeta_{2,z}), \nonumber\\
(\rho_{f})_{34}&=&(\rho_{f}^*)_{43}=\frac{1}{4u}\,(t_d-t_e)(t_d^*+t_e^*)
(\zeta_{1,z}-\zeta_{2,z}-i\zeta_{1,x}\zeta_{2,y}+i\zeta_{1,y}\zeta_{2,x}), \nonumber\\
(\rho_{f})_{44}&=&\frac{1}{4u}\,|t_d+t_e|^2
(1-\zeta_{1,x}\zeta_{2,x}-\zeta_{1,y}\zeta_{2,y}-\zeta_{1,z}\zeta_{2,z}), 
\end{eqnarray}
with
$$
u=|t_d|^2+|t_e|^2-(1+{\bm\zeta}_1{\bm\zeta}_2){\rm Re}(t_dt_e^*).
$$
For obtaining the density matrix~(\ref{rho_f_norm_avg}), we note that according to Eqs.~(\ref{rho_f_unnorm_1}) and~(\ref{out_spin_density_matrix_pure}) the unnormalized density matrices $\hat{\tilde\rho}_f(\pm{\bm\zeta}_1,\pm{\bm\zeta}_2)$ and $\hat{\tilde\rho}_f(\pm{\bm\zeta}_1,\mp{\bm\zeta}_2)$ in Eq.~(\ref{rho_f_unnorm_avg}) depend linearly on the components of the unit spin-polarization vectors ${\bm\zeta}_1$ and ${\bm\zeta}_2$. This makes the statistical averaging in  Eq.~(\ref{rho_f_unnorm_avg}) straightforward, and as a result we obtain the same expressions as in Eq.~(\ref{out_spin_density_matrix_pure}), but with ${\bm\zeta}_1$ and ${\bm\zeta}_2$ replaced by ${\bf P}_1$ and ${\bf P}_2$, respectively.

For the purpose of calculating the entanglement measures such as concurrence and entanglement of formation it is convenient to transform from Bell's basis to the conventional spin basis
\begin{equation}
|1\uparrow\rangle\otimes|2\uparrow\rangle=
\begin{pmatrix}
1\\0\\0\\0
\end{pmatrix}, \quad
|1\uparrow\rangle\otimes|2\downarrow\rangle=
\begin{pmatrix}
0\\1\\0\\0
\end{pmatrix}, \quad
|1\downarrow\rangle\otimes|2\uparrow\rangle=
\begin{pmatrix}
0\\0\\1\\0
\end{pmatrix}, \quad
|1\downarrow\rangle\otimes|2\downarrow\rangle=
\begin{pmatrix}
0\\0\\0\\1
\end{pmatrix}.
\end{equation}
According to the relations~(\ref{Bell_states}), the corresponding unitary-transformation matrix $\hat{U}$, $\hat{\rho}_f\to\hat{U}\hat{\rho}_f\hat{U}^\dag$, is given by
\begin{equation}
\hat{U}=\frac{1}{\sqrt{2}}\begin{pmatrix}
1 & 1 & 0 & 0 \\
0 & 0 & 1 & 1 \\
0 & 0 & 1 & -1 \\
1 & -1 & 0 & 0
\end{pmatrix}.
\end{equation}
%

%

%

\end{document}